\documentclass[a4paper,12pt,twocolumn]{article}  
\pdfoutput=1
\usepackage[T1]{fontenc} 
\usepackage{graphicx} 
\usepackage{bbold}
\usepackage{mathtools}
\usepackage{amsmath}
\numberwithin{equation}{section} 
\usepackage{amsfonts}
\usepackage{amssymb}
\usepackage{slashed} 
\usepackage{color}
\usepackage[table]{xcolor}
\usepackage{tabularx}
\usepackage{hyperref}
\usepackage{bm}
\usepackage{cite}  
\usepackage{mathrsfs}
\usepackage{framed}
\usepackage{xspace}
\usepackage[font=footnotesize,labelfont=bf]{caption}
\usepackage{physics}
\usepackage{mathtools}
\usepackage{subcaption}
\usepackage{array}
\newcolumntype{P}[1]{>{\centering\arraybackslash}p{#1}}
\newcolumntype{M}[1]{>{\centering\arraybackslash}m{#1}}
\def\eq#1{{Eq.~(\ref{#1})}}

\def\sec#1{{sec.~(\ref{#1})}}

\def\abs#1{\left| #1\right|}

\def\Im{\mbox{Im}\,}
\def\Re{\mbox{Re}\,}
\def\Tr{\mbox{Tr}\,}

\def\ie{{\it i.e.}\xspace}

\colorlet{grayline}{gray!70}
\definecolor{blueline}{rgb}{0,0.27,0.55}
\definecolor{DarkGray}{gray}{0.4}
\definecolor{Gray}{gray}{0.6}
\definecolor{oucrimsonred}{rgb}{0.6, 0.0, 0.0}
\definecolor{persianblue}{rgb}{0.11, 0.22, 0.73}
\definecolor{forestgreen}{rgb}{0.13,0.35,0.13}
 \hypersetup{colorlinks, citecolor=forestgreen, linkcolor=forestgreen, urlcolor=forestgreen}
\newcommand{\be}{\begin{equation}}
\newcommand{\ee}{\end{equation}}
\newcommand{\bea}{\begin{eqnarray}}
\newcommand{\eea}{\end{eqnarray}}
\newcommand{\nn}{\nonumber}

\newcommand*\xbar[1]{%
  \hbox{\;%
    \vbox{%
      \hrule height 0.5pt 
      \kern0.5ex
      \hbox{%
        \kern-0.25em
        \ensuremath{#1}%
        \kern-0.07em
      }%
    }%
  }%
} 
\newcommand{\com}[1]{}
\newcommand{\gsim}{\lower.7ex\hbox{$\;\stackrel{\textstyle>}{\sim}\;$}}
\newcommand{\lsim}{\lower.7ex\hbox{$\;\stackrel{\textstyle<}{\sim}\;$}} 

\newcommand{\bc}{\begin{center}}
\newcommand{\ec}{\end{center}}

\newcommand{\CC}{\operatorname{\bf C}}
\newcommand{\BB}{\operatorname{\bf B}}
\newcommand{\conc}[1]{{\cal C}\qty[#1]}

\newcommand{\hk}{{\bf \hat{k}}}
\newcommand{\hr}{{\bf \hat{r}}}
\newcommand{\hn}{{\bf \hat{n}}}
\newcommand{\hp}{{\bf \hat{p}}}

\newcommand{\lambdap}{\lambda^{\prime}}

\newcommand{\Pagne}{\ensuremath{\bar{\nu}_{\textrm{e}}}\xspace}
\newcommand{\Pagngm}{\ensuremath{\bar{\nu}_{\mu}}\xspace}
\newcommand{\Pagngt}{\ensuremath{\bar{\nu}_{\tau}}\xspace}
\newcommand{\Pai}{\ensuremath{\textrm{a}_{1}\textrm{(1260)}}\xspace}

\newcommand{\Pam}{\ensuremath{\textrm{a}_{1}^{-}}\xspace}
\newcommand{\Pamp}{\ensuremath{\textrm{a}_{1}^{\mp}}\xspace}
\newcommand{\Pap}{\ensuremath{\textrm{a}_{1}^{+}}\xspace}
\newcommand{\Papm}{\ensuremath{\textrm{a}_{1}^{\pm}}\xspace}
\newcommand{\PB}{\ensuremath{\textrm{B}}\xspace}

\newcommand{\Pem}{\ensuremath{\textrm{e}^{-}}\xspace}
\newcommand{\Pep}{\ensuremath{\textrm{e}^{+}}\xspace}
\newcommand{\Pgg}{\ensuremath{\gamma}\xspace}

\newcommand{\Pgmm}{\ensuremath{\mu}^{-}\xspace}

\newcommand{\Pgngt}{\ensuremath{\nu_{\tau}}\xspace}

\newcommand{\Pgpm}{\ensuremath{\pi^{-}}\xspace}
\newcommand{\Pgpp}{\ensuremath{\pi^{+}}\xspace}
\newcommand{\Pgppm}{\ensuremath{\pi^{\pm}}\xspace}
\newcommand{\Pgpz}{\ensuremath{\pi^{\textrm{0}}}\xspace}
\newcommand{\Pgri}{\ensuremath{\rho\textrm{(770)}}\xspace}

\newcommand{\Pgrm}{\ensuremath{\rho^{-}}\xspace}
\newcommand{\Pgrmp}{\ensuremath{\rho^{\mp}}\xspace}
\newcommand{\Pgrp}{\ensuremath{\rho^{+}}\xspace}
\newcommand{\Pgrpm}{\ensuremath{\rho^{\pm}}\xspace}
\newcommand{\Pgt}{\ensuremath{\tau}\xspace}
\newcommand{\Pgtm}{\ensuremath{\tau^{-}}\xspace}
\newcommand{\Pgtp}{\ensuremath{\tau^{+}}\xspace}

\newcommand{\PH}{\ensuremath{\textrm{H}}\xspace}

\newcommand{\Pp}{\ensuremath{\textrm{p}}\xspace}

\newcommand{\PZ}{\ensuremath{\textrm{Z}}\xspace}

\newcommand{\tauhp}{\ensuremath{\Pgt_{\textrm{h}}^{+}}\xspace}
\newcommand{\tauhm}{\ensuremath{\Pgt_{\textrm{h}}^{-}}\xspace}

\newcommand{\eV}{\ensuremath{\textrm{eV}}\xspace}

\newcommand{\GeV}{\ensuremath{\textrm{GeV}}\xspace}
\newcommand{\GeVinv}{\ensuremath{\textrm{GeV}^{-1}}\xspace}

\newcommand{\TeV}{\ensuremath{\textrm{TeV}}\xspace}
\newcommand{\micron}{\ensuremath{\mu\textrm{m}}\xspace}
\newcommand{\energy}{\ensuremath{E}\xspace}
\newcommand{\pT}{\ensuremath{p_{\textrm{T}}}\xspace}
\newcommand{\px}{\ensuremath{p_{\textrm{x}}}\xspace}
\newcommand{\py}{\ensuremath{p_{\textrm{y}}}\xspace}
\newcommand{\pz}{\ensuremath{p_{\textrm{z}}}\xspace}
\newcommand{\mass}{\ensuremath{\textrm{mass}}\xspace}
\newcommand{\dxy}{\ensuremath{d_{\textrm{xy}}}\xspace}
\newcommand{\dz}{\ensuremath{d_{\textrm{z}}}\xspace}
\newcommand{\fbinv}{\ensuremath{\textrm{fb}^{-1}}\xspace}

\newcommand{\s}{\ensuremath{s}\xspace}

\newcommand{\h}{\ensuremath{h}\xspace}
\newcommand{\hvec}{\ensuremath{\mathbf{h}}\xspace}
\newcommand{\hvecp}{\ensuremath{\mathbf{h}^{+}}\xspace}
\newcommand{\hvecm}{\ensuremath{\mathbf{h}^{-}}\xspace}
\newcommand{\Bvec}{\ensuremath{\operatorname{\mathbf{B}}}\xspace}
\newcommand{\Bvecp}{\ensuremath{\operatorname{\mathbf{B}}^{+}}\xspace}
\newcommand{\Bvecm}{\ensuremath{\operatorname{\mathbf{B}}^{-}}\xspace}

\newcommand{\thetap}{\ensuremath{\theta^{+}}\xspace}
\newcommand{\phip}{\ensuremath{\phi^{+}}\xspace}

\newcommand{\thetam}{\ensuremath{\theta^{-}}\xspace}
\newcommand{\phim}{\ensuremath{\phi^{-}}\xspace}

\newcommand{\Rchsh}{\ensuremath{\mathfrak{m}_{12}}\xspace}
\newcommand{\concurrence}{\ensuremath{{\cal C}[\rho]}\xspace}

\newcommand{\cosThetaCut}{\ensuremath{0.40}\xspace}
\font\beeg=cmr17 scaled 1800
\newbox\ibox
\def\versal#1{\setbox\ibox=\hbox{{\beeg #1}~}%
	    \noindent\global\hangindent=\wd\ibox\global\hangafter-2%
	    \sc\smash{\llap {\lower 14pt \box\ibox}}}

\begin{document}
\onecolumn
\thispagestyle{empty}
\begin{center}
{ \Large \color{oucrimsonred} \textbf{ 
 Probing entanglement and \\[+1pt]
 testing Bell inequality violation\\[+6pt] with $\Pep\Pem \to \Pgtp\Pgtm$ at Belle II
}}

\vspace*{1.5cm}
{\bf K. Ehat\"aht$^{a}$,}
{\bf M. Fabbrichesi$^{b}$,}
{\bf L. Marzola$^{{a}}$}
and 
 {\bf C. Veelken$^{a}$}
 

\vspace{0.5cm}
{\small 
{\it \color{DarkGray}
$^{a}$ Laboratory of High-Energy and Computational Physics, NICPB, R\"avala pst. 10, \\ 10143 Tallinn, Estonia}
\\[1mm]
{\it  \color{DarkGray} $^{b}$
INFN, Sezione di Trieste, Via Valerio 2, I-34127 Trieste, Italy}
}
\end{center}

 \vskip0.5cm
\bc
{\color{DarkGray}
\rule{0.7\textwidth}{0.5pt}}
\ec
\vskip1cm
\bc
{\bf ABSTRACT}
\ec

\vspace*{5mm}

\noindent We present a feasibility study to probe quantum entanglement and Bell inequality violation in the process $\Pep\Pem \to \Pgtp\Pgtm$ at a center-of-mass energy of $\sqrt{s} = 10.579$~\GeV. The sensitivity of the analysis is enhanced by applying a selection on the scattering angle $\vartheta$ in the $\Pgtp\Pgtm$ center-of-mass  frame. We analyze events in which both $\Pgt$ leptons decay to hadrons, using a combination of decay channels $\Pgtm \to \Pgpm\Pgngt$, $\Pgtm \to \Pgpm\Pgpz\Pgngt$, and $\Pgtm \to \Pgpm\Pgpp\Pgpm\Pgngt$. The spin orientation of the $\Pgt$ leptons in these decays is reconstructed using the polarimeter-vector method. Assuming a dataset of $200$ million $\Pgtp\Pgtm$ events and accounting for experimental resolutions, we expect the observation of quantum entanglement and Bell inequality violation by the Belle-II experiment will be possible with a significance well in excess of five standard deviations.
  
  \vskip 3cm
\bc 
{\color{DarkGray} \vbox{$\bowtie$}}
\ec


\newpage
\section{Introduction}
\label{sec:Introduction}

The most distinctive feature of quantum mechanics is the inseparable nature of states describing physical systems that have interacted in the past. The entangled states give rise to correlations between these systems that are present even after they are separated and can no longer interact. After establishing the presence of quantum entanglement, observables testing the violation of Bell inequality~\cite{Bell:1964kc} are the most interesting because they provide a direct proof of the non-local nature of quantum correlations. The Bell inequality is derived by combining the probabilities of the outcome of various measurements between two observers under the assumption of Bell locality---that is, the factorizability of these probabilities with respect to all shared resources (see, for example, \cite{Brunner:RevModPhys.86.419} for a review). Quantum mechanics does not satisfy Bell locality and the inequality can therefore be violated.

Bell inequality violation has been verified experimentally with the polarizations of low-energy (that is, few \eV) photons in~\cite{Aspect:1982fx,Weihs:1998gy}: two photons are prepared into a singlet state and their polarizations measured along different directions to verify their entanglement and the violation of Bell inequality. Many experiments have been performed to further test the inequality~\cite{Clauser:1969ny,Clauser:1974tg} and close possible loopholes~\cite{Hensen:2015ccp,Giustina:2015yza} with photons, using superconducting circuits~\cite{Storz:2023jjx}, and using atoms~\cite{PhysRevLett.119.010402}. The reader can find more details and references in two review articles~\cite{Clauser:1978ng,Genovese:2005nw}.

Though the study of entangled states and Bell inequality has been an ongoing concern in atomic and solid-state physics for many years, it is only recently that also the high-energy community has taken up in earnest the study of the subject. Collider detectors, while not designed for the probing of entanglement, turn out to be surprisingly good in performing this task, thus ushering in the possibility of many interesting new measurements as well as new tools in probing physics beyond the Standard Model. 

Entanglement with low-energy protons has been probed in Ref.~\cite{Lamehi-Rachti:1976wey} and proposed at colliders using hadronic final states in Ref.~\cite{Tornqvist:1980af}.
The higher energy probes quantum entanglement at smaller length scales~\cite{Abel:1992kz}. 
Tests in the high-energy regime of particle physics have been suggested by means of neutral kaon physics~\cite{PhysRevD.57.R1332,Benatti2000,Bertlmann:2001ea} (see also Ref.~\cite{Banerjee:2014vga}), positronium~\cite{Acin:2000cs}, flavor oscillations in neutral $\PB$-mesons~\cite{Go:2003tx}, charmonium decays~\cite{Baranov:2008zzb} and neutrino oscillations~\cite{Banerjee:2015mha}.
A discussion of some of these issues also appears in Refs.~\cite{Yongram:2013soa,Cervera-Lierta:2017tdt}. 
The interest has been revived recently after entanglement has been argued~\cite{Afik:2020onf} to be present in top-quark pair production at the LHC and it was shown~\cite{Fabbrichesi:2021npl} that Bell inequality violation is experimentally accessible in the same system. Following these works, the study of entanglement has been proposed for top quark production~\cite{Severi:2021cnj,Larkoski:2022lmv,Aguilar-Saavedra:2022uye,Afik:2022dgh,Afik:2022kwm}, hyperons~\cite{Gong:2021bcp} and gauge bosons from Higgs boson decay~\cite{Barr:2021zcp,Aguilar-Saavedra:2022wam,Ashby-Pickering:2022umy,Fabbrichesi:2023cev} and direct production~\cite{Ashby-Pickering:2022umy,Fabbrichesi:2023cev}. 
For all these particles, it is possible to study entanglement and verify the violation of Bell inequality. 
Entanglement has recently been observed in top-quark pairs produced at a center-of-mass energy of $\sqrt{s} = 13$~\TeV~\cite{ATLAS:2023fsd}.
It has also been argued~\cite{Fabbrichesi:2023idl} that Bell inequality is violated in the decays of $\PB$ mesons at LHCb and Belle II.

In this paper, we propose to probe quantum entanglement and Bell inequality violation using the process $\Pep\Pem \to \Pgtp\Pgtm$ at the Belle-II experiment, located at the SuperKEKB collider. The SuperKEKB collider delivers $\Pep\Pem$ collisions at a center-of-mass (CM) energy of $\sqrt{s} = 10.579$~\GeV. The Belle collaboration has published analyses of $\Pep\Pem \to \Pgtp\Pgtm$ production with data corresponding to an integrated luminosity of up to $921$~\fbinv, equivalent to up to $841$ million $\Pgtp\Pgtm$ events~\cite{Belle:2020lfn}, while another $175$ million events have been added with the Belle II dataset~\cite{Belle-II:2023izd}.  
The final aim of the SuperKEKB project is to collect $50$~ab$^{-1}$ of data~\cite{Akai:2018mbz,Belle-II:2018jsg}. This would result in a dataset of about $50$ billion $\Pep\Pem \to \Pgtp\Pgtm$ events.

We assess the feasibility of our proposal with a Monte Carlo (MC) study.
The study is based on $200$ million $\Pep\Pem \to \Pgtp\Pgtm$ events, which we analyze in three decay channels: $\Pgtm \to \Pgpm\Pgngt$, $\Pgtm \to \Pgpm\Pgpz\Pgngt$, and $\Pgtm \to \Pgpm\Pgpp\Pgpm\Pgngt$.
The combination of these decay channels covers about $21\%$ of $\Pgt$ pair decays. 
The detection of quantum entanglement and Bell inequality in the process $\Pep\Pem \to \Pgtp\Pgtm$ requires the measurement of $\Pgt$ spin correlations in the restframes of the $\Pgtp$ and $\Pgtm$.
Besides the availability of a large $\Pgtp\Pgtm$ dataset, our motivation for performing this study at Belle II is that the
overconstrained event kinematics and comparatively low CM energy allow for a precise reconstruction of these restframes,
which in turn allows for a precise measurement of the $\Pgt$ spin correlations in the directions transverse and longitudinal to the $\Pgt$ flight direction. The measurement of transverse and longitudinal $\Pgt$ spin correlations is important in order to distinguish quantum entanglement from local hidden-variable theories~\cite{Abel:1992kz}.

Ours is the first study of entanglement and Bell inequality violation in the process $\Pep\Pem \to \Pgtp\Pgtm$ at Belle II. Tests of entanglement and Bell inequality violation in $\Pgtp\Pgtm$ systems has previously been proposed in $\Pep\Pem$ collisions at LEP~\cite{Privitera:1991nz}, $\Pp\Pp$ collisions at the LHC~\cite{Fabbrichesi:2022ovb} and at future leptonic colliders~\cite{Altakach:2022ywa,Ma:2023yvd}.

The paper is organized as follows: in Sec.~\ref{sec:SM} we briefly summarize how the density matrix describing the polarization state of the $\tau$-pair can be computed from the amplitudes of the underlying process. Sec.~\ref{sec:EntanglementObservables} introduces the entanglement observables that we track in the following Monte Carlo analysis. In Sec.~\ref{sec:AnalysisStrategy} we propose a strategy for detecting quantum entanglement and Bell inequality violation in the data recorded by the Belle-II experiment. The details and results pertaining to the performed numerical study are described in Sec.~\ref{sec:MonteCarloStudy}, our conclusions are offered in Sec.~\ref{sec:Summary}.

\section{Tau spin correlations in the Standard Model}
\label{sec:SM}
The density matrix describing the polarization state of the bipartite system formed by the $\tau$-lepton pair can be written as
\begin{equation}
\label{eq:rho_deco}
    \rho = \frac{1}{4} \qty[
    \mathbb{1}\otimes\mathbb{1} 
    + 
    \sum_i \BB_i^+ \, \qty(\sigma_i \otimes \mathbb{1}) 
    + 
    \sum_j \BB_j^- \, \qty(\mathbb{1} \otimes \sigma_j )
    +
    \sum_{i,j} \CC_{ij} \, \qty(\sigma_i \otimes \sigma_j)
    ],
\end{equation}
where $i,j\in\{n,r,k\}$ and $\sigma_i$ are the Pauli matrices. The coefficients $\BB^\pm_i$ encode the polarization of the corresponding $\tau^\pm$ lepton, whereas the matrix $\CC_{ij}$ contains the polarization correlations. The proposed decomposition refers to a right-handed orthonormal basis $\{\hn, \hr, \hk\}$~\cite{Bernreuther:2001rq} defined in the $\tau$-pair center of mass frame as follows:
\begin{equation}
\label{eq:nrk}
    \hn = \frac{1}{\sin\vartheta}\qty(\hp \times\hk), \quad \hr = \frac{1}{\sin\vartheta}\qty(\hp-\cos\vartheta\hk) 
\end{equation}
with $\hk$ being the direction of one of the $\tau$ leptons in the center of mass frame and $\vartheta$ the angle satisfying $\hp\cdot\hk = \cos\vartheta$, with $\hp$ an arbitrary unit vector in the production plane. The quantization axis for the polarization is taken along $\hk$, so that $\sigma_k\equiv\sigma_3$. Formally, 
\begin{align}
    \BB_i^+ &= \frac{\Tr[\rho\qty(\sigma_i \otimes \mathbb{1}) ]}{4},\\
    \BB_i^- &= \frac{\Tr[\rho\qty( \mathbb{1} \otimes \sigma_i) ]}{4},\\
    \CC_{ij} &= \frac{\Tr[\rho\qty(\sigma_i \otimes \sigma_j)]}{4},
\end{align}
as implied by the properties $\Tr(\sigma_i\sigma_j)=2\delta_{ij}$ and $\Tr(\sigma_i)=0$.

The polarization density matrix can be computed from the scattering amplitude of the underlying $e^+ e^- \to \tau^+ \tau^-$ process in the following way. Consider the amplitude for the production of a fermion $\psi_\lambda$ with polarization $\lambda\in\{-\frac{1}{2},\frac{1}{2}\}$ along a given quantization direction:
\begin{equation}
    \mathcal{M}(\lambda) = \qty[\bar u_\lambda \mathcal{A}],
\end{equation}
where the symbol $\mathcal{A}$ indicates the term in the amplitude $\mathcal{M}$ multiplying the spinor $\bar u_\lambda$ and square brackets denote a contraction of spinor indices. The outgoing particle is then described by a state
\begin{equation}
    \ket{\psi} = \sum_\lambda \mathcal{M}(\lambda) \ket{u_\lambda}
\end{equation}
yielding the spinor-space density matrix
\begin{align}
   \rho_\psi 
    &= 
    \frac{\dyad{\psi}{\psi}}{\braket{\psi}} 
    = 
    \frac{\sum_{\lambda\lambdap} \qty[\bar u_\lambda\mathcal{A}]\qty[\bar u_{\lambdap}\mathcal{A}]^\dagger\,\dyad{u_\lambda}{\bar u_{\lambdap}}}{{\sum_{\lambda\lambdap}\qty[\bar u_\lambda\mathcal{A}]\qty[\bar u_{\lambdap}\mathcal{A}]^\dagger\,\braket{\bar u_{\lambdap}}{u_{\lambda}}}} 
    \\ \nn
    &=
     \frac{\sum_{\lambda\lambdap}\qty[\bar u_\lambda\mathcal{A}]\qty[\bar u_{\lambdap}\mathcal{A}]^\dagger\,\dyad{u_\lambda}{\bar u_{\lambdap}}}{2m{\sum_{\lambda}\qty[\bar u_\lambda\mathcal{A}]\qty[\bar u_{\lambda}\mathcal{A}]^\dagger}}
    =
    \frac{\sum_{\lambda\lambdap}\qty[\mathcal{A}\bar u_{\lambdap}]^\dagger \qty[\mathcal{A}\bar u_\lambda]\,\dyad{u_\lambda}{\bar u_{\lambdap}}}
    {2m \,\abs{\mathcal{M}}^2}
    ,
\end{align}
where we made use of the orthogonality relation $\braket{\bar u_{\lambdap}}{u_{\lambda}} \equiv \qty[\bar u_{\lambdap} u_{\lambda}] = 2 m \delta_{\lambdap \lambda}$ with $m$ being the mass of the fermion and $\abs{\mathcal{M}}^2$ the squared amplitude (summed over the spin) for the production process. 

The density matrix in the polarization space can be obtained upon projection via the operators~\cite{Bouchiat:1958yui}
\begin{align}
    \label{eq:projU}
    \frac{\dyad{u_\lambda}{\bar u_{\lambdap}}}{2m}&\equiv\frac{\Pi_{\lambda\lambdap}^u}{2m}= \frac{1}{4m}\qty(\slashed{p}+m)\qty(\delta_{\lambda \lambdap}+\gamma_5\sum_i\slashed{s_i}\sigma^i_{\lambda \lambdap})
    \\
    \label{eq:projV}
    \frac{\dyad{v_\lambda}{\bar v_{\lambdap}}}{2m}&\equiv\frac{\Pi_{\lambda\lambdap}^v}{2m}= \frac{1}{4m}\qty(\slashed{p}-m)\qty(\delta_{\lambda \lambdap}+\gamma_5\sum_i\slashed{s_i}\sigma^i_{\lambda \lambdap})
\end{align}
where $\{s_i^\mu\}$ is a triad of space-like four-vectors, satisfying $s_i^\mu p_\mu =0$, obtained by boosting the canonical basis\footnote{In the rest frame of the fermion we have $s=(0, {\bf s})$ and
\begin{equation}
\nn
s_1 = \begin{pmatrix}
0 \\ 1 \\ 0 \\ 0    
\end{pmatrix}, \qquad 
s_2 = \begin{pmatrix}
0 \\ 0 \\ 1 \\ 0    
\end{pmatrix}, \qquad
s_3 = \begin{pmatrix}
0 \\ 0 \\ 0 \\ 1    
\end{pmatrix}.  
\end{equation}} of the spin four-vector $s$ to the frame where the fermion has four-momentum $p$. By acting with the first projector on the spinor-space density matrix we then obtain   
\begin{equation}
    \rho_{\lambda \lambdap}=\qty[\frac{\Pi_{\lambda\lambdap}^u}{2m} \rho_\psi] = \frac{\qty[\mathcal{A}\mathcal{A}^\dagger\Pi_{\lambda\lambdap}^u]}
    {\abs{\mathcal{M}}^2} = \frac{1}{2}\qty(\mathbb{1}+\sum_{i}\expval{s_i}\sigma^i),
\end{equation}
where $i\in\{n,r,k\}$ and $\expval{X}$ is the ensemble average of the quantity $X$. The computation of the density matrix for an anti-fermion $\bar\psi_\lambda$ proceeds analogously with the replacement of the projection operator in~\eq{eq:projU} by that of~\eq{eq:projV}. The generalization to processes yielding more than one fermion in the final state is straightforward and recovers~\eq{eq:rho_deco} for the pair production case. In particular we have $\BB^\pm_i=\expval{s_i^\pm}$, $\CC_{ij}=\expval{s_i^+\, s_j^-}$, with $s^+_i$ and $s^-_i$ being the spin vector of the anti-fermion and fermion, respectively. 

\begin{figure}[h]
\centering
\includegraphics[scale=1]{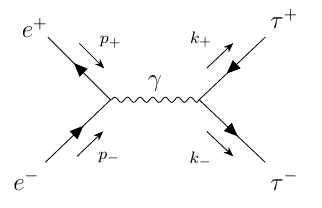}
\caption{The tree-level Feynman diagram for the considered process. We omit $\PZ$ and $\PH$ contributions as they are negligible at $\sqrt{s} = 10.579$~\GeV. }
\label{fig:diag}
\end{figure}

The amplitude of the process we are interested in can be obtained from the diagram in Fig.~\ref{fig:diag}, yielding 
\begin{equation}
\mathcal{M}=
-\frac{e^2}{s} \left[\bar{v}(p_+)\gamma^\mu u(p_-)\right]\left[ \bar u(k_-)\gamma_\mu v(k_+)\right]
\end{equation}
where $s=(p_+ + p_-)^2$ and the squared amplitude 
\begin{equation}
    \abs{\mathcal{M}}^2=\frac{4 e^4}{s}  \left(4 \, m_{\Pgt}^2\sin^2\vartheta + s\left(\cos^2\vartheta + 1\right) \right)
\end{equation}
where we neglected the mass of the electron and summed over the spin of all the involved fermions. In the $\tau$-pair center of mass frame, our conventions set $\hk$ along the $k_+$ momentum, $\hp$ along $p_+$ and identify the angle $\vartheta$ in Eq.~(\ref{eq:nrk}) with the scattering angle included by these momenta. 

Computing the polarization density matrix of the $\Pgt$-pair we find 
\begin{equation}
\rho = r_0
{\begin{pmatrix}
 s \left(\cos^2 \vartheta +1\right) & -i \sqrt{s}m_{\tau } \sin 2\vartheta  & -i \sqrt{s}m_{\tau } \sin 2\vartheta  & -s \sin^2 \vartheta 
 \\
 i \sqrt{s}m_{\tau } \sin 2\vartheta  & 4 m_{\tau }^2\sin^2 \vartheta   & 4 m_{\tau }^2\sin^2 \vartheta   & i \sqrt{s}m_{\tau } \sin 2\vartheta  
 \\
 i \sqrt{s}m_{\tau } \sin 2\vartheta  & 4 m_{\tau }^2\sin^2 \vartheta  & 4 m_{\tau }^2\sin^2 \vartheta  & i \sqrt{s}m_{\tau } \sin 2\vartheta 
 \\
 -s \sin^2 \vartheta & -i \sqrt{s}m_{\tau } \sin 2\vartheta & -i \sqrt{s}m_{\tau } \sin 2\vartheta & s \left(\cos^2 \vartheta +1\right) 
\end{pmatrix}},
\end{equation}
where $r_0=1/(8  m_{\tau }^2\sin^2 \vartheta +2s (\cos^2\vartheta +1))$.

For the coefficients in \eq{eq:rho_deco} we obtain $\BB^\pm_i = 0$  and
\begin{align}
\label{eq:C}
\CC 
 &= c_{0}
\begin{pmatrix}
 \left(4 m_{\tau }^2 - s\right) \sin^2\vartheta
 & 
 0 
 & 
 0 
 \\
 0 
 & 
 \left( 4 m_{\tau }^2 + s\right)\sin^2\vartheta
 & 
 4 m_{\Pgt}\sqrt{s} \sin\theta\cos\vartheta
 \\
 0 & 
 4 m_{\Pgt}\sqrt{s} \sin\vartheta  \cos\vartheta
 & 
 -4 m_{\Pgt}^2 \sin^2\vartheta + s\left(\cos^{2}\vartheta + 1\right)
 \\
\end{pmatrix} \,
\end{align}
where $c_{0} = 1/\left(4 m_{\Pgt}^2 \sin^2\vartheta + s\left(\cos^{2}\vartheta + 1\right)\right)$.
Averaging over the angular distribution of the two $\Pgt$ leptons yields
\begin{equation}
   \expval{\CC}=
\left(
\begin{array}{ccc}
-0.419875 & 0 & 0\\
0 & 0.526708 & 0\\
0 & 0 & 0.893167 
\end{array}
\right)
\label{eq:C_sm}
\end{equation}
for $\sqrt{s} = 10.579$ GeV and $m_\tau=1.777$ GeV. 

\eq{eq:C_sm} shows that the parameter $D=\Tr(\expval{\CC})/3=1/3$ does not signal the presence of entanglement as it relies on an average. Averaging, in general, dilutes the effect of quantum correlations.

\section{Entanglement observables}
\label{sec:EntanglementObservables}
On general grounds, a bipartite state is called \textit{separable} if its density matrix can be written as a convex combination of product states:
\begin{equation}
\rho_{\text{sep}}=\sum_{i,j}p_{ij}\,\rho_i^{A}\otimes\rho_j^{B}, \quad\text{with}\quad p_{ij}>0 \quad\text{and}\quad\sum_{i,j}p_{ij}=1 \, ,
\end{equation}
where the labels $A$ and $B$ denote the two composing subsystems. By definition, a system is called \textit{entangled} if it is \textit{not} separable.

Quantifying the entanglement content of a bipartite system is generally a complicated task because the possible decompositions into pure state pose an optimization problem for the chosen entanglement measure or monotone. Fortunately, algebraic solutions are available for simpler systems, for instance, for those composed of two qubits. In order to assess the presence of entanglement in the $\tau$-pair polarization state we track the \textit{concurrence} $\conc{\rho}$~\cite{Bennett:1996gf,Horodecki:2009zz, Wootters:1997id}, an entanglement monotone which for a bipartite qubit system can be quantified as 
\begin{equation}
    \label{eq:concurrence}
    \conc{\rho}=\max\qty{0, \lambda_1-\lambda_2-\lambda_3-\lambda_4}\in \qty[0,1],
\end{equation}
where $\lambda_i$ are the eigenvalues, in decreasing order, of the matrix
\be
R=\sqrt{\sqrt{\rho} \tilde\rho\sqrt{\rho}}, \quad{\text{with}}\quad \tilde\rho = \qty(\sigma_2\otimes\sigma_2)\rho^*\qty(\sigma_2\otimes\sigma_2).
\ee
Non-vanishing values of the concurrence witness the presence of entanglement and a value of 1 indicates a maximally entangled state. At the tree level we find
\begin{equation}
    \conc{\rho}=\frac{\left(s - 4 m_{\Pgt}^2\right) \sin^2\vartheta}{4 m_{\Pgt}^{2} \sin^2\vartheta + s \left(\cos^2\vartheta + 1\right)}
\end{equation}
for the process $\Pep\Pem \to \Pgtp\Pgtm$.
Entanglement vanishes at the kinematic threshold because the conservation of angular momentum, in absence of an orbital component, forces the complete classical correlation of the $\tau$-pair spins and their state into a separable one.

\begin{figure}[t!]
\begin{center}
\includegraphics[width=4.5in]{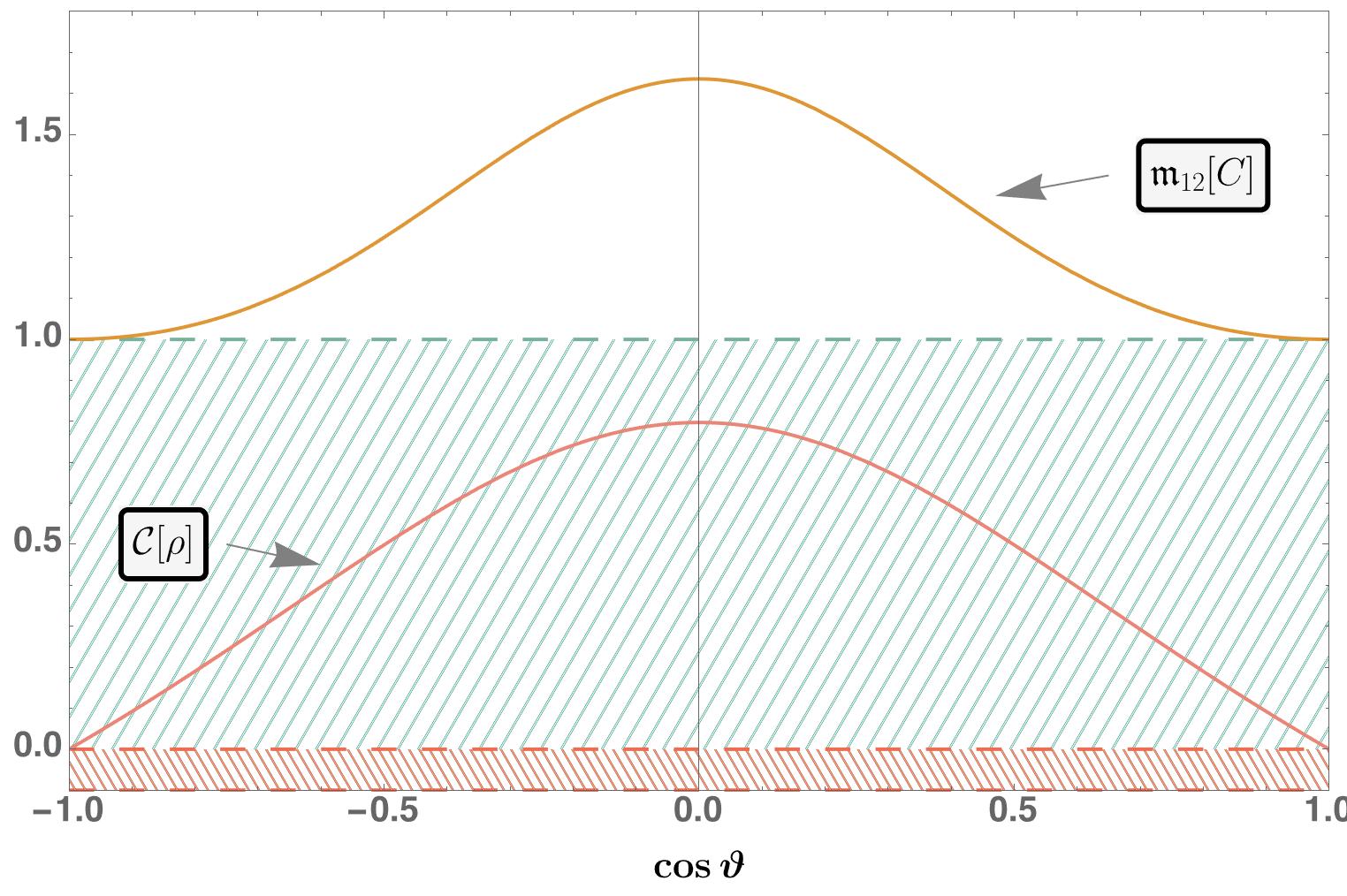}
\caption{\small The observables $\Rchsh\qty[\CC]$ and $\concurrence$ as a function of the scattering angle $\vartheta$ (defined by the directions of the incoming positron and outgoing $\tau^+$ in the $\tau$-pair rest frame) at $\sqrt{s}=10.579$ GeV. The solid red and yellow lines represent the Standard Model expectation for these observables. Entanglement is present if $\concurrence>0$ (above the area hatched in red), while the generalized Bell inequalities are violated for $\Rchsh\qty[\CC]>1$ (above the area hatched in green). For both observables, the central region (for small  $\cos \vartheta$) is where the largest values are to be found.
\label{fig:m12} 
}
\end{center}
\end{figure}

\begin{figure}[t!]
\begin{center}
\includegraphics[width=4.5in]{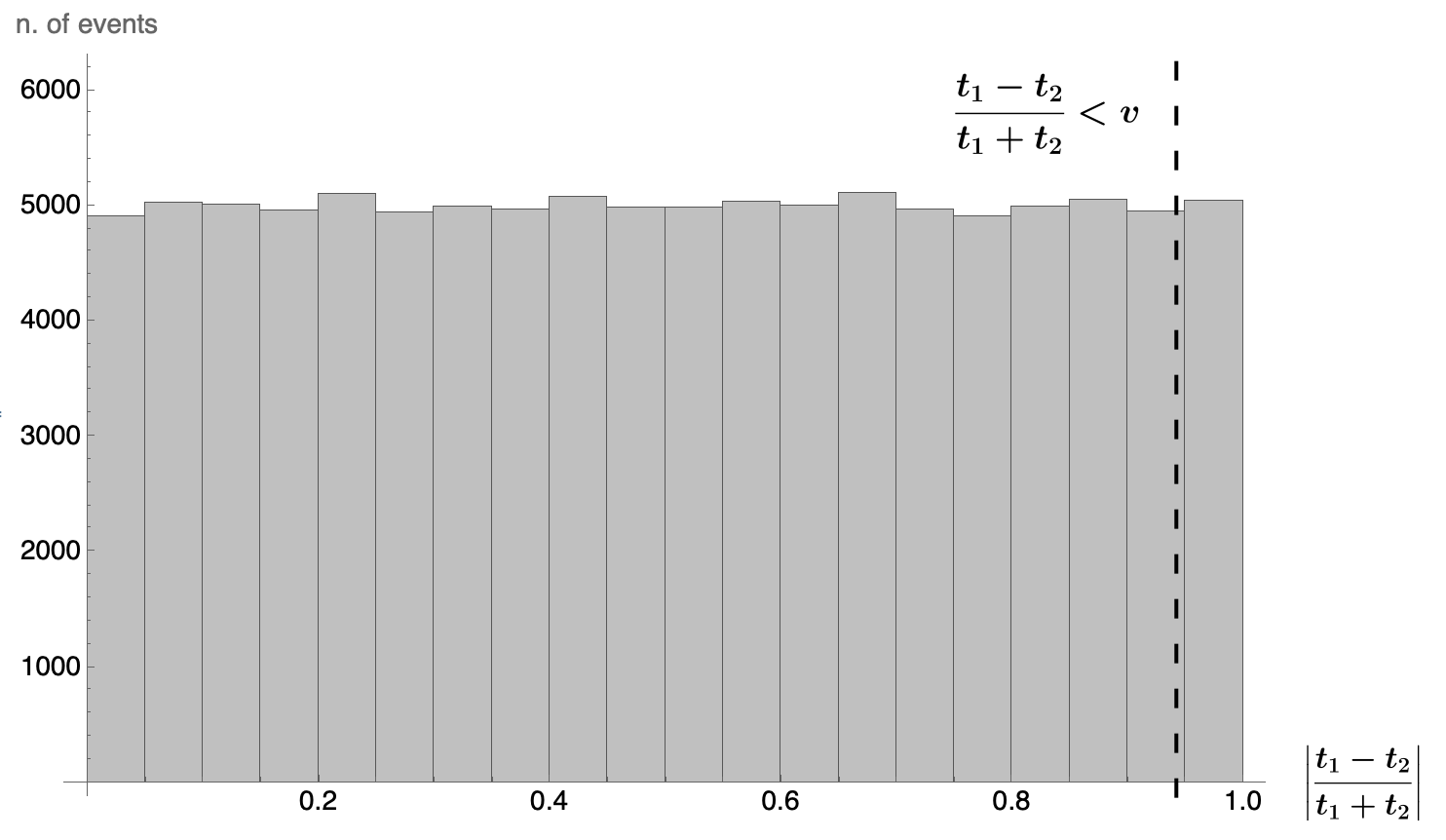}
\caption{\small Histogram of the number of events as a function of the ratio $|t_1-t_2|/(t_1+t_2)$ between the difference and the sum of the decay times of the two taus. The events have been generated by $10^5$ pseudo-experiments in which the decay times are randomly varied within an exponential distribution. The black-dashed vertical line distinguishes events separated by a time-like interval (to the right of the line) from those that are space-like separated (to the left of the line).
\label{fig:ratio} 
}
\end{center}
\end{figure}

The genuine quantum correlations that entangle the polarization states of the $\tau$ lepton can also be used to discriminate between quantum mechanics and alternative local stochastic classical theories relying on hidden variables~\cite{Clauser_1978}. This is the idea behind the so-called Bell inequalities~\cite{Bell:1964kc}, which bound the expectation value of a suitable operator under the hypothesis that the involved correlators are \textit{local}, i.e. that they factorize according to the rules of probability~\cite{Privitera:1991nz}. For the bipartite qubit system at hand, a useful test is encoded in the following inequality~\cite{Clauser:1969ny}   
\begin{equation}
\label{eq:gen_Bell}
    \abs{\hat{n}_1\cdot\CC\cdot\qty(\hat{n}_2-\hat{n}_4)+\hat{n}_3\cdot\CC\cdot\qty(\hat{n}_2+\hat{n}_4)}\leq 2,
\end{equation}
with $\hat{n}_i$ being four unit vectors indicating the directions along which the spins of the two leptons can be measured. The upper bound is respected by  correlations stemming from local theories but \textit{can} be violated within quantum mechanics if the state of interest is entangled. In order to detect the violation of this generalized Bell inequality it is necessary to maximize the effect through a suitable choice of the four measurement directions. The procedure can be bypassed by introducing the operator $\Rchsh\qty[\CC]$~\cite{HORODECKI1995340}, defined as    
\begin{equation}
    \Rchsh\qty[\CC] = m_1+m_2
\label{eq:Rchsh}
\end{equation}
where $m_1\geq m_2\geq m_3$ are the eigenvalues of the positive semi-definite matrix $M=\CC^T\CC$. If and only if $\Rchsh\qty[\CC]>1$, then the bound in \eq{eq:gen_Bell} is violated and local hidden-variable theories can be ruled out.   

With the results in \eq{eq:C} we find
\begin{equation}
   \Rchsh\qty[\CC] = 1 + \left(\frac{\left(s - 4 m_{\Pgt}^2\right) \sin^2\vartheta}{4 m_{\Pgt}^2 \sin^2\vartheta + s \left(\cos^2\vartheta + 1\right)}\right)^{2} \, ,
\end{equation}
which we plot in Fig.~\ref{fig:m12} as a function of the scattering angle. The figure shows that the violation of the bound in \eq{eq:gen_Bell} becomes easier to detect when one selects events in which the $\Pgt$ pair is produced in direction transverse to the beam axis.

In relation to the proposed Bell test, we remark that the relative velocity $v$ with which the $\Pgt$-pair flies apart is sufficiently large to create, at the times $t_1$ and $t_2$ of decay, a space-like separation
\be 
\frac{|t_1-t_2|\, c}{(t_1+t_2)\, v} < 1\, ,
\ee
for more than $95\%$ of the $\Pgt$ pairs (see Fig.\ref{fig:ratio}).
The separation prevents local interactions (as those arising through the exchange of photons between the charged taus) and ensures that the locality loophole~\cite{Bell:1987hh} is closed. The selection of these events could be implemented with a suitable cut on the relative momentum of the two particles. However, given the amount of available data and the small fraction of pairs rejected by the cut, this refinement would not affect the significance of the proposed Bell test.

\section{Measurement of tau spin correlations}
\label{sec:AnalysisStrategy}

The measurement of the $\Pgt$ spin correlation matrix $\CC$ is based on the spin-dependent differential cross section $d\sigma$ for the process $\Pep\Pem \to \Pgtp\Pgtm$, which is given by~\cite{Jadach:1990mz}:
\begin{equation}
d\sigma = \vert\mathcal{\bar{M}_{\textrm{p}}}\vert^{2} \, \left( 1 - b^{+}_{\mu} \, \s_{+}^{\mu} - b^{-}_{\nu} \, \s_{-}^{\nu} + c_{\mu\nu} \, \s_{+}^{\mu} \, \s_{-}^{\nu} \right) \, d\textrm{Lips} \, ,
\label{eq:diffXS1}
\end{equation}
where $\vert\mathcal{\bar{M}_{\textrm{p}}}\vert^{2}$ denotes the spin-averaged matrix element for the process $\Pep\Pem \to \Pgtp\Pgtm$ and $d\textrm{Lips}$ is the Lorentz invariant phase-space measure.
The symbols $\s_{+}^{\mu}$ and $\s_{-}^{\mu}$ refer to the spin of the $\Pgtp$ and of the $\Pgtm$.
Combining Eq.~(\ref{eq:diffXS1}) with the expression for the differential decay rate of the $\Pgt$, given by~\cite{Jadach:1990mz}:
\begin{equation}
d\Gamma = \frac{1}{2 \, m_{\Pgt}} \, \vert\mathcal{\bar{M}_{\textrm{d}}}\vert^{2} \, \left( 1 + \h_{\mu} \, \s^{\mu} \right) d\textrm{Lips} \, ,
\label{eq:diffDecayRate}
\end{equation}
one obtains the relation:
\begin{equation}
d\sigma = \vert\mathcal{\bar{M}_{\textrm{p}}}\vert^{2} \, \vert\mathcal{\bar{M}_{\textrm{d}}}\vert^{2} \, \vert\mathcal{\bar{M}'_{\textrm{d}}}\vert^{2} \, \left( 1 - b^{+}_{\mu} \, \h_{+}^{\mu} - b^{-}_{\nu} \, \h_{-}^{\nu} + c_{\mu\nu} \, \h_{+}^{\mu} \, \h_{-}^{\nu} \right) \, d\textrm{Lips} \, .
\label{eq:diffXS2}
\end{equation}
The symbols $\vert\mathcal{\bar{M}_{\textrm{d}}}\vert^{2}$ and $\vert\mathcal{\bar{M}'_{\textrm{d}}}\vert^{2}$ refer to the spin-averaged matrix elements for the decay of the $\Pgtp$ and $\Pgtm$ and $\h_{+}^{\mu}$ and $\h_{-}^{\nu}$ denote the polarimeter vectors of the $\Pgtp$ and $\Pgtm$, respectively.
The polarimeter vectors provide a handle to measure the orientation of the $\Pgtp$ and $\Pgtm$ spins.
The relation between the polarimeter vector and the $\Pgt$ spin orientation is given by Eq.~(\ref{eq:diffDecayRate}).

For the three decay channels $\Pgtm \to \Pgpm\Pgngt$, $\Pgtm \to \Pgpm\Pgpz\Pgngt$, and $\Pgtm \to \Pgpm\Pgpp\Pgpm\Pgngt$ considered in this paper, the polarimeter vector is a function of the momenta of the charged and neutral pions produced in the $\Pgt$ decay.
For the decay channels $\Pgtm \to \Pgpm\Pgngt$ and $\Pgtm \to \Pgpm\Pgpz\Pgngt$ the polarimeter vector can be computed analytically, and we use the expressions given by Eqs.~(3.25) and~(3.39) of Ref.~\cite{Jadach:1990mz}. For the decay channel $\Pgtm \to \Pgpm\Pgpp\Pgpm\Pgngt$ it is not possible to derive analytic expressions for the polarimeter vector and we instead use the algorithm of Ref.~\cite{Cherepanov:2023wfp} to compute $h_{+}^{\mu}$ and $h_{-}^{\nu}$ numerically.
The decays $\Pgtm \to \Pgpm\Pgpz\Pgngt$ and $\Pgtm \to \Pgpm\Pgpp\Pgpm\Pgngt$ proceed via intermediate $\Pgri$ and $\Pai$ resonances. We hence refer to these decay channels as $\Pgpp$, $\Pgrp$, and $\Pap$ for the $\Pgtp$ and as $\Pgpm$, $\Pgrm$, and $\Pam$ for the $\Pgtm$.

It has been shown that all hadronic $\Pgt$ decay channels provide the same sensitivity, or ``$\Pgt$ spin analyzing power'', if the  charged and neutral pions produced in the $\Pgt$ decays can be reconstructed and measured with negligible experimental resolution~\cite{Kuhn:1995nn}.
In contrast, the spin analyzing power of the leptonic decay channels $\Pgtm \to \Pem\Pagne\Pgngt$ and $\Pgtm \to \Pgmm\Pagngm\Pgngt$ is limited to about $40\%$ compared to hadronic $\Pgt$ decays~\cite{Davier:1992nw}.
For this reason, we focus on hadronic $\Pgt$ decays in this paper.

The polarimeter vectors $\h_{+}^{\mu}$ and $\h_{-}^{\nu}$ need to be computed in the restframes of the $\Pgtp$ and $\Pgtm$.
The restframes are determined by reconstructing the full event kinematics, including the momenta of the two neutrinos produced in the $\Pgt$ decays, as detailed in the next Section.
In the $\Pgtp$ and $\Pgtm$ restframes, the timelike component of the polarimeter vector vanishes: $h^{0} = 0$.
Eq.~(\ref{eq:diffXS2}) thus reduces to:
\begin{equation}
d\sigma = \vert\mathcal{\bar{M}_{\textrm{p}}}\vert^{2} \, \vert\mathcal{\bar{M}_{\textrm{d}}}\vert^{2} \, \vert\mathcal{\bar{M}'_{\textrm{d}}}\vert^{2} \, \left( 1 + \Bvecp \cdot \hvecp + \Bvecm \cdot \hvecm + \hvecp \cdot \CC \cdot \;\hvecm \right) \, d\textrm{Lips} \, .
\label{eq:diffXS3}
\end{equation}

Using this relation, we determine the polarizations $\Bvecp$ and $\Bvecm$ 
and the spin correlation matrix $\CC$ by an unbinned maximum-likelihood (ML) fit~\cite{Rossi:2018}.
The likelihood function is given by:
\begin{equation}
\mathcal{L} = \prod_{i} \, \left( 1 + \Bvecp \cdot \hvecp_{i} + \Bvecm \cdot \hvecm_{i} + \hvecp_{i} \cdot \CC \cdot \;\hvecm_{i} \right) \, .
\label{eq:lf}
\end{equation}
In Eq.~(\ref{eq:lf}), the subscript $i$ refers to the events $i$ in the $\Pep\Pem \to \Pgtp\Pgtm$ event sample and the product extends over all events in this sample. 
The $15$ parameters of the fit: the $3$ elements of the polarization vector $\Bvecp$ of the $\Pgtp$, the $3$ elements of the polarization vector $\Bvecm$ of the $\Pgtm$, and the $9$ elements of the spin correlation matrix $\CC$ are determined by a numerical maximization of the likelihood function $\mathcal{L}$ with respect to these parameters.
The parameters are expressed in the $\{\hn, \hr, \hk\}$ coordinate system defined in Section~\ref{sec:SM}.
The maximization is performed numerically, using the program MINUIT~\cite{James:1975dr}.
Alternative procedures for determining $\Bvecp$, $\Bvecm$, and $\CC$ are compared to the ML-fit method in Section~\ref{sec:appendixB} of the Appendix.

Eq.~(\ref{eq:diffXS3}) holds for a fixed value of the scattering angle $\vartheta$. We have checked that the maximization of the likelihood function yields an unbiased estimate of the spin correlation matrix $\CC$ when Eq.~(\ref{eq:C}) is integrated over intervals in $\vartheta$.
As an example, we give in Eq.~(\ref{eq:C_pipi}) the spin correlation matrix computed for a sample of $\Pep\Pem \to \Pgtp\Pgtm$ events, produced by MC simulation as detailed in Section~\ref{sec:MonteCarloStudy}:
\begin{equation}
\label{eq:C_pipi}
\CC =
\begin{pmatrix}
 -0.4129 \pm 0.0033 & -0.0014 \pm 0.0040 & \;\;\;0.0008 \pm 0.0036 \\
  \;\;\;0.0007 \pm 0.0029 &  \;\;\;0.5273 \pm 0.0032 & \;\;\;0.0024 \pm 0.0030 \\
  \;\;\;0.0024 \pm 0.0031 &  \;\;\;0.0030 \pm 0.0032 & \;\;\;0.8829 \pm 0.0028 
\end{pmatrix}
\end{equation}
The events considered in the computation were selected in the decay channel $\Pgpp\Pgpm$ within the range $0 \leq \vartheta \leq \pi$ and were analyzed on MC-truth level. No selection criteria (acceptance cuts) are applied on the charged and neutral pions produced in the $\Pgt$ decays. All $9$ elements of the matrix agree with the SM expectation, given by Eq.~(\ref{eq:C_sm}), within the quoted statistical uncertainties.

Once the spin correlation matrix $\CC$ is determined, we compute the observables $\conc{\rho}$ and $\Rchsh\qty[\CC]$ using Eqs.~(\ref{eq:concurrence}) and~(\ref{eq:Rchsh}), in order to test for entanglement and Belle inequality violation. 
The elements of the density matrix $\rho$ are given by $\Bvecp$, $\Bvecm$, and $\CC$ through Eq.~(\ref{eq:rho_deco}).

Fig.~\ref{fig:m12} demonstrates that the feasibility to detect quantum entanglement and Belle inequality violation increases if one selects events in which the $\Pgt$ leptons are produced in direction transverse to the beam axis, \ie with $\vartheta \approx \pi/2$. We perform an optimization of a selection on $\vartheta$, with the aim of maximizing the significances $\conc{\rho}/\delta\conc{\rho}$ and $(\Rchsh\qty[\CC] - 1)/\delta\Rchsh\qty[\CC]$. The results of this optimization will be presented in the next Section.

\section{Monte Carlo study}
\label{sec:MonteCarloStudy}

A sample of $200$ million $\Pep\Pem \to \Pgtp\Pgtm$ MC events was generated with the program MadGraph\_aMCatNLO v2.9.16~\cite{Alwall:2014hca}, using leading-order matrix elements.
The program PYTHIA v8.306~\cite{Bierlich:2022pfr} is used for the modeling of parton showers, hadronization processes, and $\Pgt$ decays.
All $\Pgt$ decay channels are included in the simulation.
The events are analyzed on MC-truth level and after taking realistic experimental resolutions into account.

Instead of performing a full simulation of the Belle II detector~\cite{Belle-II:2010dht} based on GEANT4~\cite{GEANT4:2002zbu},
we simulate the experimental resolution by randomly varying (``smearing'') the position of the primary event vertex, the four-vectors of the charged and neutral particles produced in the $\Pgt$ decays, and the longitudinal ($\dxy$) and transverse ($\dz$) impact parameters of tracks. In case of $\Pgt$ decays into three charged pions, we also smear the position of the $\Pgt$ decay vertex.
The $z$ axis is defined as the direction of the electron beam. 
For the $\Pgt$ decay channels considered in this paper, only the resolutions for charged pions ($\Pgppm$) and for photons ($\Pgg$) are relevant. The latter originate from neutral pion ($\Pgpz$) decays.
The resolutions are taken from Ref.~\cite{Belle-II:2010dht} and are summarized in Table~\ref{tab:ExperimentalResolution}.
For the $\Pgt$ decay vertex, we assume a resolution of $500$~\micron in direction parallel to the $\Pgt$ flight direction and $10$~\micron in each of the two perpendicular directions.
The smeared values are obtained by randomly sampling from a Gaussian distribution with mean equal to the true value and width equal to the experimental resolution given in the table.
The symbol $\pT$ refers to the momentum in direction transverse to the beam axis, and the symbols $\theta$ and $\phi$ denote the polar and azimuthal angles with respect to this axis.
The resolution on the $\pT$ of charged pions is parametrized by $\sigma_{\pT} = \pT \, ( c_{0} \, \pT \oplus c_{1}/\beta )$, where $\beta = \sqrt{1 - (m/E)^{2}}$ is the charged pions' velocity in units of the speed of light.
The resolution on the energy $\energy$ of photons is parametrized by $\sigma_{\energy} = \energy \, ( c_{0}/\energy \oplus c_{1}/\sqrt[4]{\energy} \oplus c_{2} )$. The coefficients $c_{i}$ are given in Table~\ref{tab:ExperimentalResolution}.
The symbol $\oplus$ indicates that contributions to the resolutions are added in quadrature.
Angular resolutions are given in units of radians.
The angular resolution for $\Pgppm$ represents our estimate.
The angular resolution for $\Pgg$ improves proportional to the square-root of the photons' energy.
For the resolutions on the transverse and longitudinal impact parameters, which typically vary with $\pT$ and $\theta$ of the track, we have taken averages and rounded the values to one significant digit.
The energy spread of the beam electrons and the effect of beamstrahlung is simulated by varying the constraint on the energy of the $\Pgtp\Pgtm$ system and of its momentum in beam direction by $0.1$~\GeV~\cite{Akai:2018mbz},
and by varying its momentum in direction transverse to the beam axis by $0.01$~\GeV.

\begin{table*}
\centering
\begin{tabular}{c c}
{\begin{tabular}{M{2.0cm}|M{3.5cm}}
\multicolumn{2}{c}{Charged hadrons} \\
Quantity & Resolution \\
\hline
$\pT$: $c_{0}$ & $1 \times 10^{-3}$~\GeVinv \\
$\pT$: $c_{1}$ & $3 \times 10^{-3}$ \\
$\theta$ & $10^{-3}$ \\
$\phi$   & $10^{-3}$ \\
$\dxy$   & $10$~\micron \\
$\dz$    & $20$~\micron \\
\end{tabular}}
&
{\begin{tabular}{M{2.0cm}|M{3.5cm}}
\multicolumn{2}{c}{Photons} \\
Quantity & Resolution \\
\hline
$\energy$: $c_{0}$ & $2 \times 10^{-3}$~\GeV \\
$\energy$: $c_{1}$ & $1.6 \times 10^{-2}~\sqrt[4]{\GeV}$ \\
$\energy$: $c_{2}$ & $1.2 \times 10^{-2}$ \\
$\theta$  & $4 \times 10^{-3}/\sqrt{\energy [\GeV]}$ \\
$\phi$    & $4 \times 10^{-3}/\sqrt{\energy [\GeV]}$ \\
\end{tabular}}
\\[+2.5cm]
{\begin{tabular}{M{2.5cm}|M{2.5cm}}
\multicolumn{2}{c}{Event vertex} \\
Quantity & Resolution \\
\hline
$x$ & $10$~\micron \\
$y$ & $10$~\micron \\
$z$ & $20$~\micron \\
\end{tabular}}
&
{\begin{tabular}{M{2.5cm}|M{2.5cm}}
\multicolumn{2}{c}{$\Pgtp\Pgtm$ system} \\
Quantity & Resolution \\
\hline
$\px$   & $0.01$~\GeV \\
$\py$   & $0.01$~\GeV \\
$\pz$   & $0.1$~\GeV \\
$\mass$ & $0.1$~\GeV \\
\end{tabular}}
\end{tabular}
\caption{Experimental resolutions used in the MC study.}
\label{tab:ExperimentalResolution}
\end{table*}

The $\Pgppm$ and $\Pgg$ produced in the $\Pgt$ decays are required to pass selection criteria, which ensure that the particles can be well identified and their momenta be well reconstructed in the Belle-II detector.
Charged pions are required to have a transverse momentum $\pT > 0.1$~\GeV, while photons are required to have an energy $E > 0.1$~\GeV.
Both are required to be within the geometric acceptance of the central drift chamber: $17 < \theta < 150^{\circ}$.
The selection criteria are taken from Ref.~\cite{Belle:2020lfn}.
We refer to them as acceptance cuts.

The full kinematics of each event, including the momenta of the two neutrinos produced in the $\Pgt$ decays, is reconstructed using a two-step procedure.
In the first step, we determine approximate values of the $\Pgt$ lepton momenta by solving a set of analytic equations.
The approximate values are then used as starting point for a kinematic fit (KF), which is executed in the second step.

The first step is based on the formalism introduced in Appendix C of Ref.~\cite{Altakach:2022ywa} and has been extended to the case of $\Pgt$ decay channels other than $\Pgtp\Pgtm \to \Pgpp\Pagngt\Pgpm\Pgngt$.
Details of the extended formalism are given in Section~\ref{sec:appendixA} of the appendix.

The KF is based on the work presented in Refs.~\cite{Avery:kinfit,Sauerland:1358627,Cherepanov:2018npf}.
The number of fitted parameters totals $17$: the position of the primary event vertex (3); the components $\px$ and $\py$ of the momenta of the neutrinos produced in the decay of the $\Pgtp$ and $\Pgtm$ (4); the position of the decay vertices of $\Pgtp$ and $\Pgtm$ (6); and the $\energy$, $\px$, $\py$, and $\mass$ components of the $\Pgtp\Pgtm$ system (4).
For the uncertainties on these parameters, the KF uses the values given in Table~\ref{tab:ExperimentalResolution}.
We use the symbols $\tauhp$ and $\tauhm$ to refer to the system of $\Pgppm$ and $\Pgg$ produced in the decays of the $\Pgtp$ and $\Pgtm$, respectively. We assume the uncertainties on the $\Pgppm$ and $\Pgg$ momenta to be negligible and thus do not include the $\tauhp$ and $\tauhm$ momenta as parameters in the fit.
For $\Pgt$ leptons that decay into $\Pgppm$ or $\Pgrpm$, we follow the approach referred to as ``huge error method'' in Ref.~\cite{Avery:hugeerror} to allow the fit to freely vary the position of the $\Pgt$ decay vertex along the direction of the charged pion's track.
The $\energy$ and $\pz$ components of the neutrino four-vector are computed analytically, using the $\Pgt$ and neutrino mass constraints.
All other constraints are represented by Lagrange multipliers in the KF. In total there are $8$ such constraints:
$4$ ``parallelism'' constraints of the type described in Section 4.1.3.3 of Ref.~\cite{Sauerland:1358627} and $4$ constraints that demand that the sum of $\tauhp + \Pgngt + \tauhm + \Pagngt$ four-vectors equals the four-vector of the $\Pep\Pem$ initial state. 

\begin{figure*}[ht!]
\centering
\includegraphics[width=0.48\textwidth]{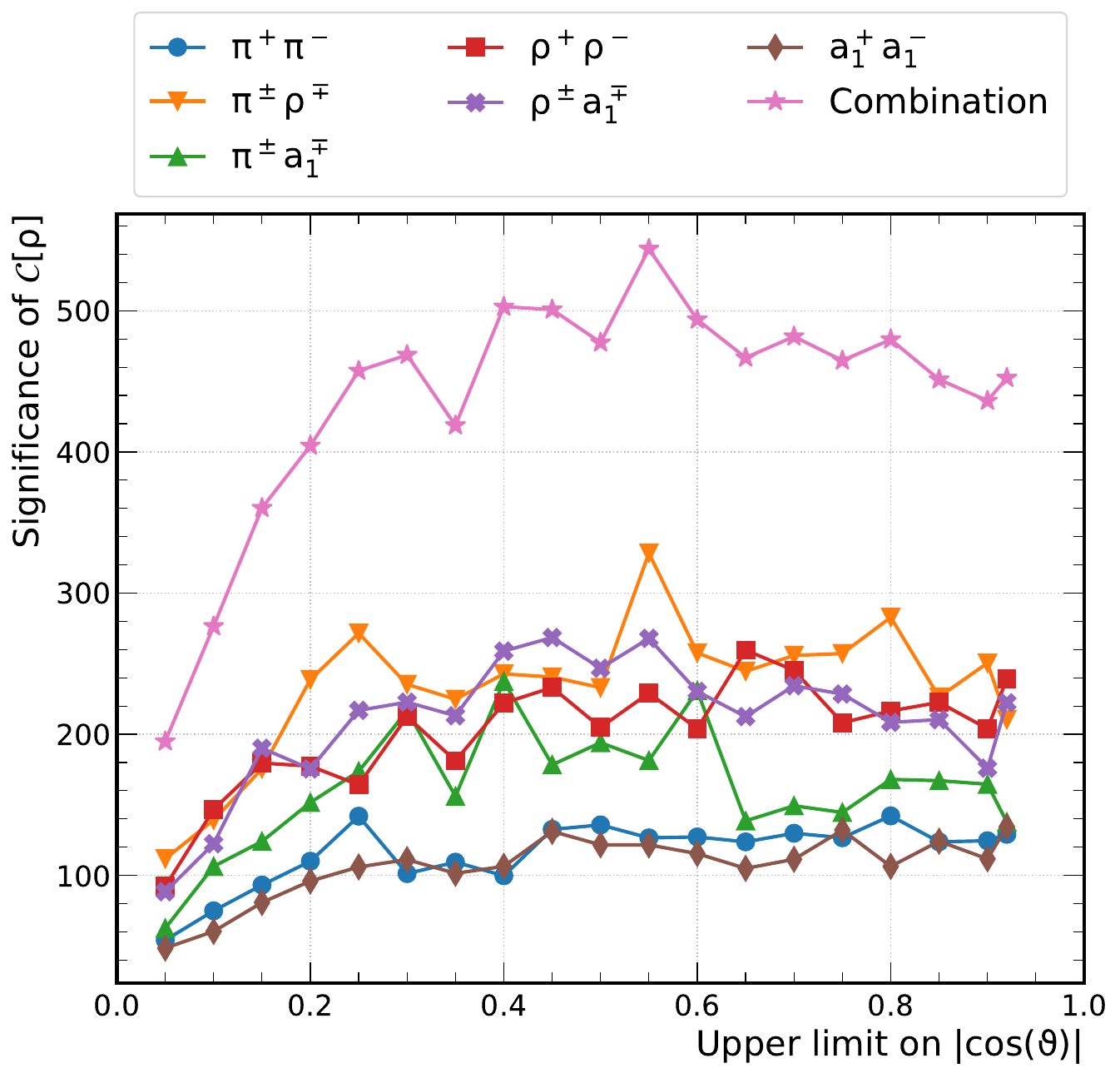}
\includegraphics[width=0.48\textwidth]{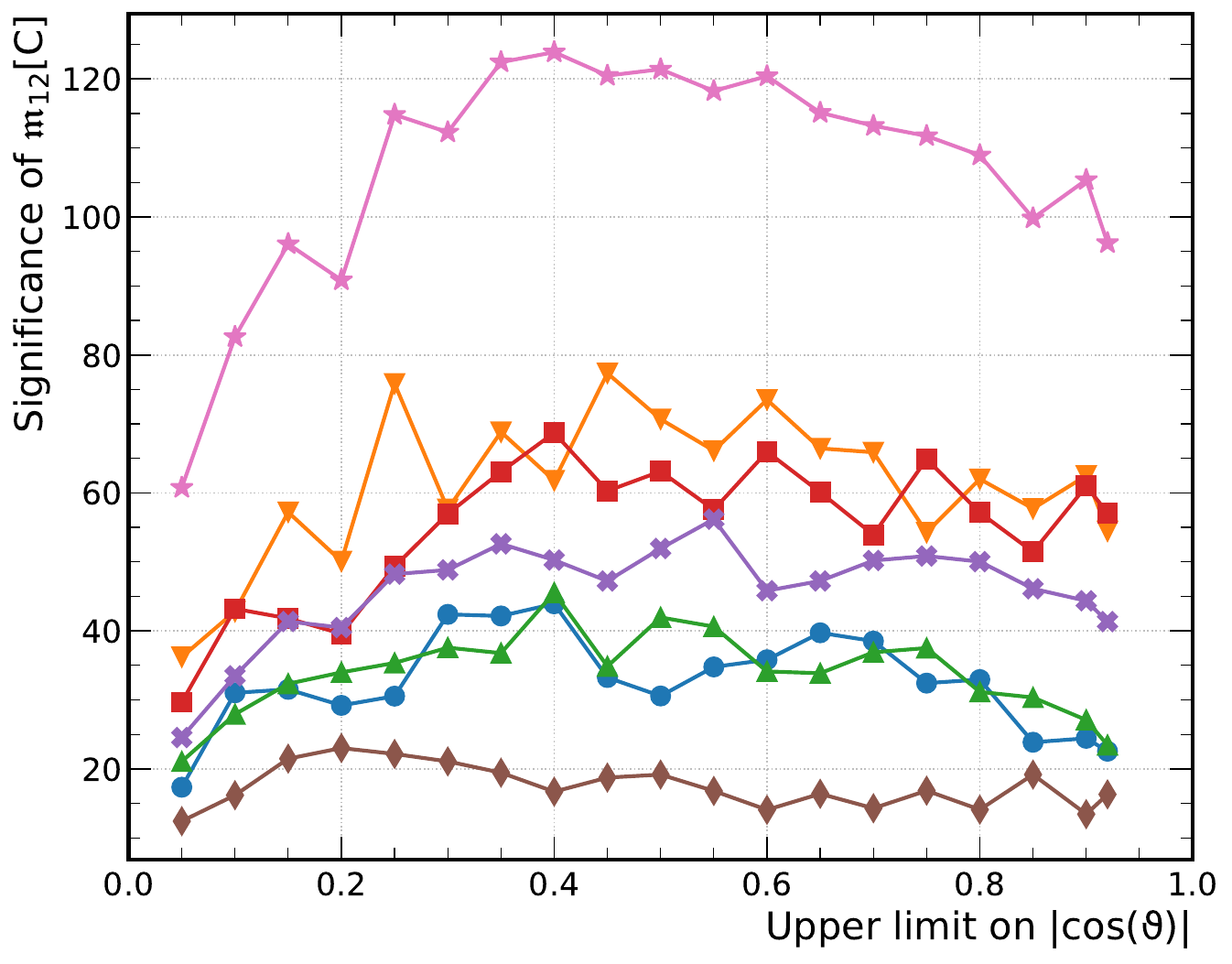}
\includegraphics[width=0.48\textwidth]{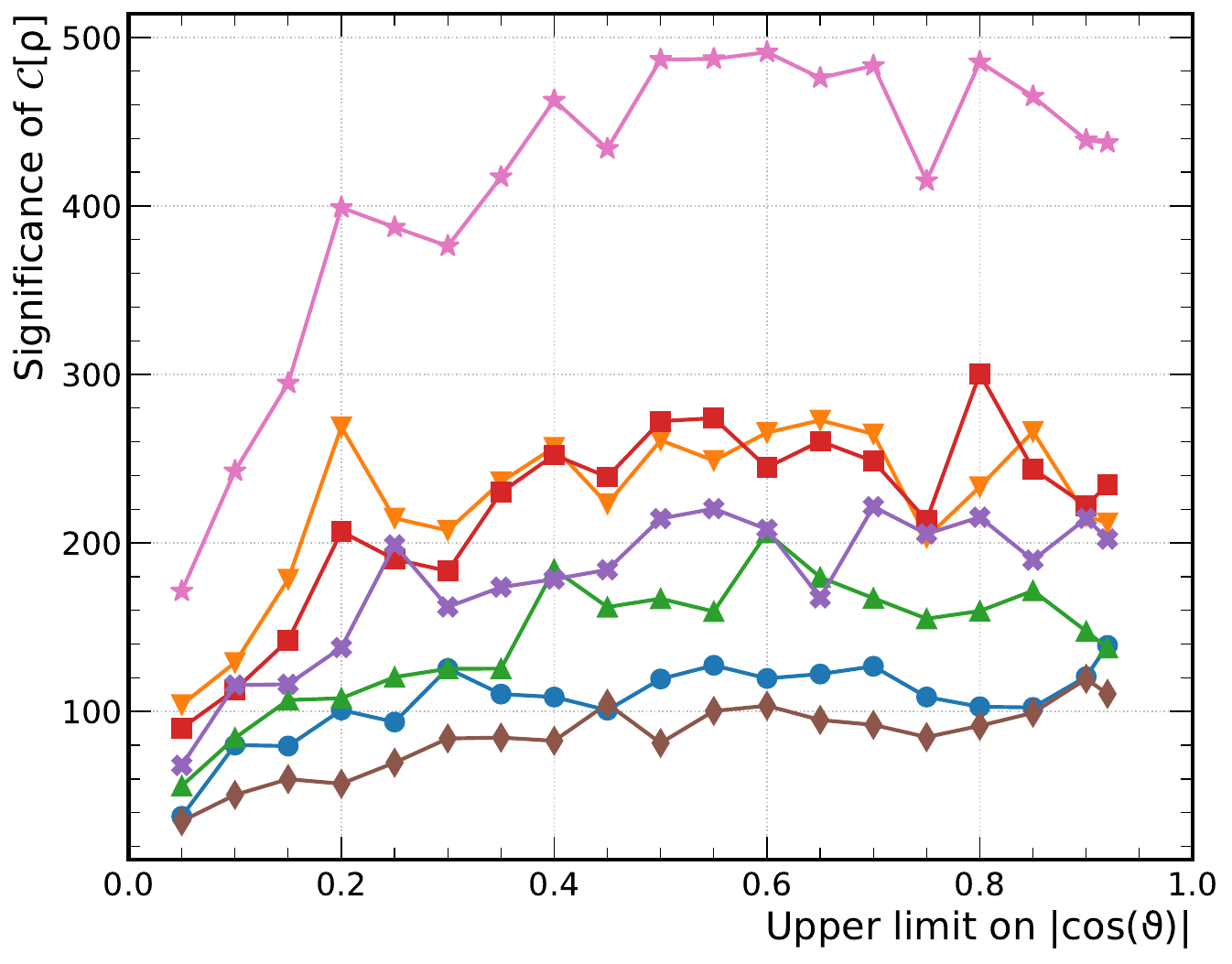}
\includegraphics[width=0.48\textwidth]{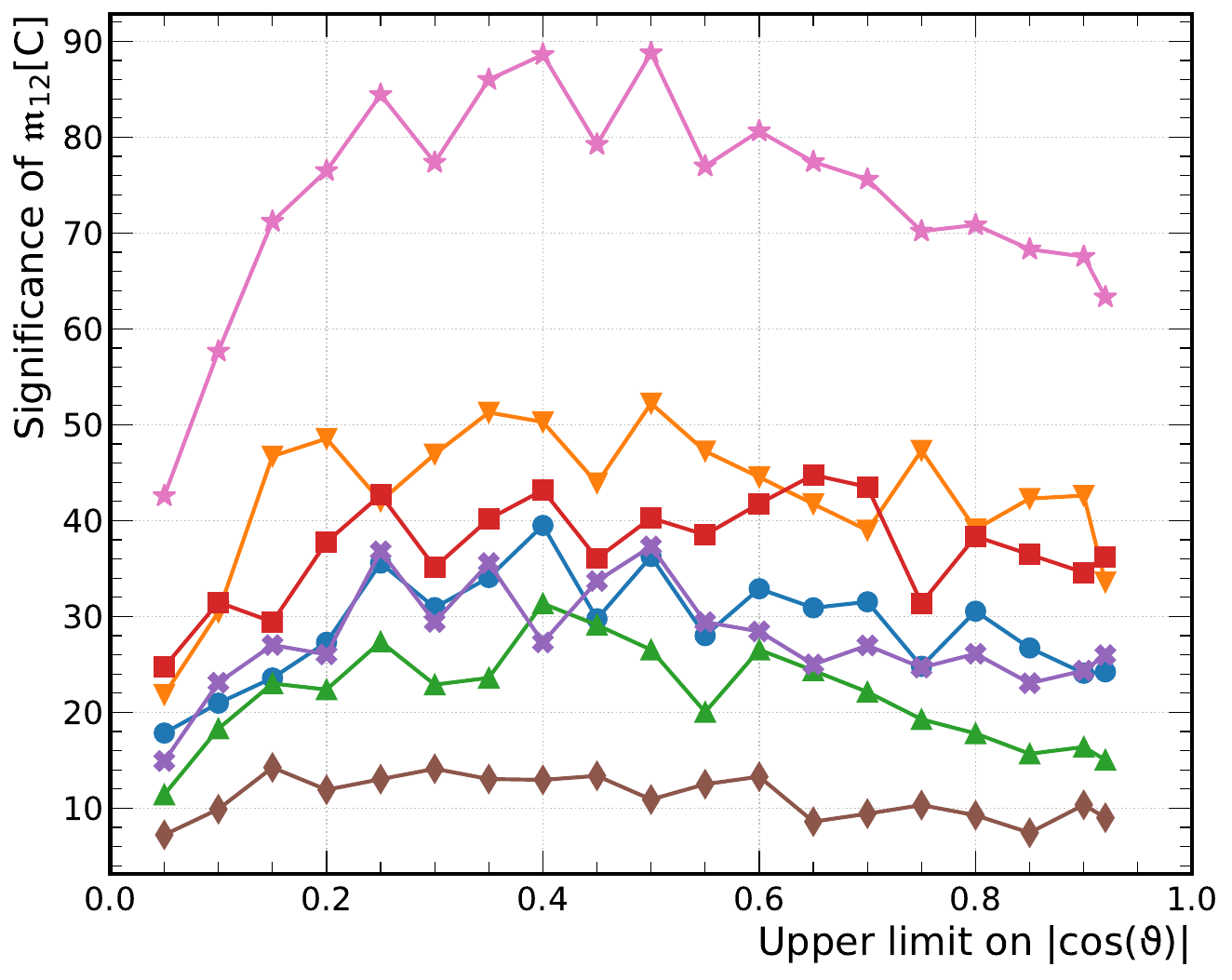}
\caption{\small
  Significances $\conc{\rho}/\delta\conc{\rho}$ (left) and $(\Rchsh\qty[\CC] - 1)/\delta\Rchsh\qty[\CC]$ (right) as function of the upper limit imposed on $\vert\cos(\vartheta)\vert$. The significances are given for the decay channels $\Pgpp\Pgpm$, $\Pgppm\Pgrmp$, $\Pgppm\Pamp$, $\Pgrpm\Pgrmp$, $\Pgrpm\Pamp$, and $\Pap\Pam$ individually and for their combination.
  The events are analyzed on MC-truth level (top) and with experimental resolutions taken into account (bottom).
  No acceptance cuts are applied on the $\Pgppm$ and $\Pgg$ produced in the $\Pgt$ decays.
}
\label{fig:theta_scan}
\end{figure*}

For the optimization of the selection on $\vert\cos(\vartheta)\vert < x$, we perform a scan of the significances $\conc{\rho}/\delta\conc{\rho}$ and $(\Rchsh\qty[\CC] - 1)/\delta\Rchsh\qty[\CC]$ as function of the upper limit $x$,
where the symbols $\delta\conc{\rho}$ and $\delta\Rchsh\qty[\CC]$ refer to the statistical uncertainties on the observables $\conc{\rho}$ and $\Rchsh\qty[\CC]$.
We vary $x$ within the range $0$ to $1$ in steps of $0.05$.
The result of the scan is illustrated in Fig.~\ref{fig:theta_scan} for the cases that the event kinematics and polarimeter vectors are taken from MC-truth level and for the case that they are reconstructed by the KF after smearing the events by the experimental resolutions.
A good compromise between maximizing the effect of entanglement and Bell inequality violation on the one hand and maintaining a high-statistics event sample on the other hand is achieved for the selection $\vert\cos(\vartheta)\vert < \cosThetaCut$.
Maintaining a high-statistics event sample reduces the uncertainties $\delta\conc{\rho}$ and $\delta\Rchsh\qty[\CC]$.

In the figure, one can see that all decay channels contribute in a meaningful way to the sensitivity for detecting quantum entanglement and Bell inequality violation. The decay channels $\Pgrp\Pgrm$ and $\Pgppm\Pgrmp$ contribute the most, reflecting their higher branching fractions. The significance of the combination is computed by adding the significances of individual decay channels in quadrature.
The significances $\conc{\rho}/\delta\conc{\rho}$ and $(\Rchsh\qty[\CC] - 1)/\delta\Rchsh\qty[\CC]$ are reduced by about $10$ and $30\%$, respectively, when the events are smeared by the experimental resolutions and reconstructed by the KF, compared to the sensitivity obtained at MC-truth level.

Values and uncertainties on the observables $\conc{\rho}$ and $\Rchsh\qty[\CC]$ for events that pass the selection $\vert\cos(\vartheta)\vert < \cosThetaCut$ on the scattering angle $\vartheta$ are given in Tables~\ref{tab:entanglObs_alldecaychannels_mctruth} and~\ref{tab:entanglObs_alldecaychannels_smeared}. Table~\ref{tab:entanglObs_alldecaychannels_mctruth} gives the results obtained when events are analyzed at MC-truth level and Table~\ref{tab:entanglObs_alldecaychannels_smeared} those obtained after smearing the events by the experimental resolutions and reconstructing the event kinematics by the KF.

\begin{table}[ht!]
\centering
\begin{tabular}{c|r@{$ \,\,\pm\,\, $}rr@{$ \,\,\pm\,\, $}r}
Decay channel & \multicolumn{2}{c}{$\conc{\rho}$} & \multicolumn{2}{c}{$\Rchsh\qty[\CC]$} \\
\hline
$\Pgpp\Pgpm$   & $0.7079$ & $0.0071$ & $1.483$ & $0.011$ \\ 
$\Pgppm\Pgrmp$ & $0.7113$ & $0.0029$ & $1.482$ & $0.008$ \\ 
$\Pgppm\Pamp$  & $0.6762$ & $0.0028$ & $1.388$ & $0.009$ \\ 
$\Pgrp\Pgrm$   & $0.7111$ & $0.0032$ & $1.495$ & $0.007$ \\ 
$\Pgrpm\Pamp$  & $0.6798$ & $0.0026$ & $1.402$ & $0.008$ \\ 
$\Pap\Pam$     & $0.6386$ & $0.0060$ & $1.294$ & $0.018$ \\ 
All channels   & $0.6905$ & $0.0014$ & $1.444$ & $0.004$ \\ 
\end{tabular}
\caption{
  Observables $\conc{\rho}$ and $\Rchsh\qty[\CC]$ measured in individual decay channels and for the combination of all six channels,
  for events that pass the acceptance cuts and the selection $\vert\cos(\vartheta)\vert < \cosThetaCut$.
  Events are analyzed on MC-truth level.
}
\label{tab:entanglObs_alldecaychannels_mctruth}
\end{table}

\begin{table}[ht!]
\centering
\begin{tabular}{c|r@{$ \,\,\pm\,\, $}rr@{$ \,\,\pm\,\, $}r}
Decay channel & \multicolumn{2}{c}{$\conc{\rho}$} & \multicolumn{2}{c}{$\Rchsh\qty[\CC]$} \\
\hline
$\Pgpp\Pgpm$   & $0.6722$ & $0.0062$ & $1.463$ & $0.012$ \\ 
$\Pgppm\Pgrmp$ & $0.6658$ & $0.0026$ & $1.361$ & $0.007$ \\ 
$\Pgppm\Pamp$  & $0.6370$ & $0.0035$ & $1.298$ & $0.009$ \\ 
$\Pgrp\Pgrm$   & $0.6524$ & $0.0026$ & $1.326$ & $0.008$ \\ 
$\Pgrpm\Pamp$  & $0.6181$ & $0.0035$ & $1.264$ & $0.010$ \\ 
$\Pap\Pam$     & $0.6062$ & $0.0073$ & $1.229$ & $0.018$ \\ 
All channels   & $0.6475$ & $0.0014$ & $1.331$ & $0.004$ \\ 
\end{tabular}
\caption{
  Observables $\conc{\rho}$ and $\Rchsh\qty[\CC]$ measured in individual decay channels and for the combination of all six channels,
  for events that pass the acceptance cuts and the selection $\vert\cos(\vartheta)\vert < \cosThetaCut$.
  Events are reconstructed by the KF after smearing them by the experimental resolutions given in Table~\ref{tab:ExperimentalResolution}.
}
\label{tab:entanglObs_alldecaychannels_smeared}
\end{table}

The uncertainties on $\Bvecp$, $\Bvecm$, and $\CC$ as well as on the observables $\conc{\rho}$ and $\Rchsh\qty[\CC]$ are computed by bootstrapping~\cite{bootstrapping}: A set of $N_{\textrm{toy}} = 100$ toy datasets are constructed from the original sample. The events in each toy datasets are drawn randomly from the original sample, such that the number of events in each toy dataset equals $N$. 
The bootstrap samples may contain the same event exactly once, multiple times, or not at all.
The probability $P(n)$ for a certain event to be contained $n$ times in the toy dataset is given by the Poisson distribution, $P(n) = (\lambda^{n} \, e^{-\lambda})/n!$ with $\lambda = 1/N$.
For each toy dataset, we compute the spin correlation matrix $\CC$ by maximizing the likelihood function $\mathcal{L}$ given by Eq.~(\ref{eq:lf}).
The statistical uncertainty on the element $\CC_{ij}$ is then computed by sorting the $N_{\textrm{toy}}$ values of this element, which we obtained by the bootstrapping procedure, and taking half the difference between the $84$ and $16\%$ quantiles.
Statistical uncertainties on the observables $\conc{\rho}$ and $\Rchsh\qty[\CC]$ are estimated by taking the set of $N_{\textrm{toy}}$ spin correlation matrices $\CC$, computing $\conc{\rho}$ and $\Rchsh\qty[\CC]$ for each matrix using Eqs.~(\ref{eq:concurrence}) and~(\ref{eq:Rchsh}), and then taking half the difference between the $84$ and $16\%$ quantiles for these observables. 

The values and uncertainties for the combination of decay channels in Tables~\ref{tab:entanglObs_alldecaychannels_mctruth} and~\ref{tab:entanglObs_alldecaychannels_smeared} are computed by taking a weighted average of the individual decay channels $i$, with weights given by the inverse of the square of the uncertainties $\delta\conc{\rho}_{i}$ and $\delta\Rchsh\qty[\CC]_{i}$ expected for channel $i$.
For the combination of all six decay channels, we expect that a measurement of $\Pgt$ spin correlations in the process $\Pep\Pem \to \Pgtp\Pgtm$ at Belle II will allow to 
observe entanglement and Bell inequality violation with significances well in excess of five standard deviations (s.d.). The numerical values of the significances amount to $463$ s.d. in case of entanglement and $87$ s.d. in case of Bell inequality violation.

The significances computed based on the numbers given in Tables~\ref{tab:entanglObs_alldecaychannels_mctruth} and~\ref{tab:entanglObs_alldecaychannels_smeared} are about $10\%$ lower compares to those shown in Fig.~\ref{fig:theta_scan}.
The difference is due to the acceptance cuts. Events passing the selection $\vert\cos(\vartheta)\vert < \cosThetaCut$ pass the acceptance cuts with an efficiency that varies between $52$ and $95\%$, depending on the $\Pgt$ decay channel.
The effect of these efficiencies is to increase the uncertainties $\delta\conc{\rho}$ and $\delta\Rchsh\qty[\CC]$.
The efficiency is the lowest for the decay channel $\Pgrp\Pgrm$ and the highest for the decay channel $\Pgpp\Pgpm$.
We have checked that the acceptance cuts do not introduce a bias on the $\Pgt$ spin correlation. The values of $\conc{\rho}$ and $\Rchsh\qty[\CC]$ obtained for events passing the selection $\vert\cos(\vartheta)\vert < \cosThetaCut$ change only marginally, by about $1\%$, when the acceptance cuts are applied.

As can be seen by comparing Tables~\ref{tab:entanglObs_alldecaychannels_mctruth} and~\ref{tab:entanglObs_alldecaychannels_smeared}, the experimental resolutions increase the uncertainties $\delta\conc{\rho}$ and $\delta\Rchsh\qty[\CC]$ by a small amount and also reduce the measured values and thus the significances by a few percent. As the focus of this study is the detection of entanglement and Bell inequality violation and not the precise measurement of $\conc{\rho}$ and $\Rchsh\qty[\CC]$, we do not make an attempt to mitigate this effect by accounting for the experimental resolutions in the likelihood function $\mathcal{L}$ given by Eq.~(\ref{eq:lf}) or compensate for the effect via calibration.

We advise the reader not to take the quoted values of the significances literally.
We quote their numerical values solely for the purpose of substantiating our expectation that an observation of entanglement and Bell inequality violation with a significance well in excess of five s.d. is highly likely after all experimental effects, including effects not considered in our Monte Carlo study, are taken into account.
The following three effects will degrade the sensitivity somewhat in a realistic experiment: the presence of non-Gaussian tails in the experimental resolutions, the presence of backgrounds, and systematic uncertainties. 
Of the three, the presence of backgrounds will probably have the most sizeable effect.

Based on Fig.~1 of Ref.~\cite{Belle-II:2023izd}, we expect the dominant background to arise from misreconstruction of the $\Pgt$ decay channel. The figure shows that the $\Pgt$ decay channel gets misreconstructed in about $15\%$ of $\Pep\Pem \to \Pgtp\Pgtm$ events at Belle II, while in the remaining $85\%$ of events the $\Pgt$ decay channel is reconstructed correctly. Backgrounds arising from the process $\Pep\Pem \to \textrm{q}\bar{\textrm{q}}$ and from other sources are small in comparison.
The misreconstruction of the $\Pgt$ decay channel may happen if, for example, the $\Pgg$ produced in $\Pgpz$ decays fail to get reconstructed due to detection inefficiencies, are outside of the geometric acceptance of the electromagnetic calorimeter, or have energies below the threshold of $0.1$~\GeV.
In case the $\Pgt$ decay channel does get misreconstructed, two things happen: Because the polarimeter vector of the $\Pgt$ depends on the momenta of the particles produced in the $\Pgt$ decay in a way that is specific to each $\Pgt$ decay channel, the polarimeter vector will be computed in the wrong way. Besides, the wrong four-vectors will be used in that computation.
Unfortunately, the full Belle-II detector simulation based on GEANT4~\cite{GEANT4:2002zbu} is necessary to study the misreconstruction of the $\Pgt$ decay channel in detail.

Concerning the effect of systematic uncertainties, we point out that the experimental resolutions considered in our MC study reduce the significances for observing entanglement and Bell inequality violation by rather moderate amounts of $10$ and $30\%$, respectively.
Since systematic uncertainties refer to uncertainties on the experimental resolutions and on the background, we expect their effect to be comparable in size to the effect of the experimental resolutions and of the background.
If systematic uncertainties constitute a limiting factor to the sensitivity of the analysis, we expect that their effect can be mitigated by including suitable control regions into the ML fit or by auxiliary measurements, taking advantage of the large dataset of $\Pep\Pem \to \Pgtp\Pgtm$ events recorded by Belle II.

We expect non-Gaussian tails to affect the numerical values of the significances, but not to alter our conclusion that an observation of entanglement and Bell inequality violation with a significance well in excess of five s.d. is highly likely once the Belle II collaboration reproduces our study with their data.

In summary, we are confident that the presence of non-Gaussian tails in the experimental resolutions, backgrounds, and systematic uncertainties disregarded in our analysis will not prevent the observation of entanglement and Bell inequality violation in the process $\Pep\Pem \to \Pgtp\Pgtm$ at Belle II.

\section{Summary}
\label{sec:Summary}

We have studied the prospects for testing QM by probing entanglement and Bell inequality violation in the process $\Pep\Pem \to \Pgtp\Pgtm$ at Belle II. We expect that a dataset of $200$ million $\Pep\Pem \to \Pgtp\Pgtm$ events will be sufficient to observe quantum entanglement and Bell inequality violation with a significance well in excess of five standard deviations after full detector effects, backgrounds, and systematic uncertainties are taken into account. A dataset of this size has already been recorded by Belle II.

Our study is based on the analysis of six decay channels: $\Pgpp\Pgpm$, $\Pgppm\Pgrmp$, $\Pgppm\Pamp$, $\Pgrp\Pgrm$, $\Pgrpm\Pamp$, and $\Pap\Pam$. Compared to analyzing only the decay channel $\Pgpp\Pgpm$, the channel most prominantly studied in the context of $\Pgt$ spin measurements in the literature, the combination of all six decay channels improves the significance for detecting entanglement by more than a factor of four and the significance for detecting Bell inequality violation by more than a factor of two. The inclusion of the decay channels $\Pgppm\Pgrmp$, $\Pgppm\Pamp$, $\Pgrp\Pgrm$, $\Pgrpm\Pamp$, and $\Pap\Pam$ into the analysis will be possible if the charged and neutral pions produced in the $\Pgt$ decays can be reconstructed with high efficiency and purity.

We encourage the Belle II collaboration to reproduce this study with their full detector simulation and their data.

\section*{Acknowledgments}
L.M. is supported by the Estonian Research Council grant PRG356. K.E. and C.V. are supported by the Estonian Research Council grant PRG445. We thank Kristjan Kannike for useful discussions, the Galileo Galilei Institute for Theoretical Physics for the hospitality, and the INFN for partial support during the completion of this work.

\newpage
\appendix

\section{Analytic equations for kinematic reconstruction}
\label{sec:appendixA}

The formalism introduced in Appendix C of Ref.~\cite{Altakach:2022ywa} yields approximate values for the eight unknown components of the $\Pgtp$ and $\Pgtm$ four-vectors by applying eight constraints and solving the resulting system of linear equations. Two constraints refer to the mass of the $\Pgt$ lepton four-vector and two require the neutrinos to be massless. The remaining four constraints are obtained by demanding the four-vector of the $\Pgtp\Pgtm$ system to be equal to the initial state of the $\Pep\Pem$ collision: $p_{\Pgt\Pgt} = (E_{\Pep} + E_{\Pem}, 0, 0, E_{\Pem} - E_{\Pep})$, where $E_{\Pep}$ and $E_{\Pem}$ refer to the nominal energies of the $\Pep$ and $\Pem$ beams, respectively, and the $z$-axis points in direction of the electron beam.
We have extended the formalism to the case of arbitrary hadronic $\Pgt$ decay channels,
obtaining the following relations, which we use in lieu of the equations given in Appendix C of Ref.~\cite{Altakach:2022ywa}.

We start by parametrizing the four-vectors of the $\Pgtp$ and $\Pgtm$, denoted by the symbols $p_{\Pgtp}$ and $p_{\Pgtm}$, by:
\begin{equation}
p_{\Pgt^{\pm}}^{\mu} = \frac{1 \mp a}{2} \, p_{\Pgt\Pgt} \pm \frac{b}{2} \, p_{\h^{+}}^{\mu} \mp \frac{c}{2} \, p_{\h^{-}}^{\mu} \pm d \, q_{\mu} \, ,
\end{equation}
where
\begin{equation}
q^{\mu} = \frac{1}{s} \, \epsilon^{\mu\nu\rho\sigma} \, p_{\Pgt\Pgt}^{\nu} \, p_{\h^{+}}^{\rho} \, p_{\h^{-}}^{\sigma} \, ,
\end{equation}
and the symbols $a$, $b$, $c$, and $d$ represent four coefficients, which are to be determined.
The first three of the coefficients, $a$, $b$, and $c$, are obtained as solutions to the equation:
\begin{equation}
\left(
\begin{array}{c}
 a \\
 b \\
 c \\ 
\end{array}
\right) = [M]^{-1} \cdot {\bf \Lambda} \, ,
\end{equation}
where
\begin{equation}
[M] = \left(
\begin{array}{ccc}
-x & m_{\h^{+}}^{2} & -z \\
 y & -z & m_{\h^{-}}^{2} \\
 s & -x & y \\ 
\end{array}
\right) \quad \textrm{and} \quad 
\quad {\bf \Lambda} = \left(
\begin{array}{c}
 m_{\Pgt}^{2} + m_{\h^{+}}^{2} - x \\
 m_{\Pgt}^{2} + m_{\h^{-}}^{2} - y \\
 0 \\ 
\end{array}
\right) \, .
\label{eq:step1_M}
\end{equation}
In the above equations, the symbols $p_{\h^{+}}$ and $p_{\h^{-}}$ denote the momentum of the $\tauhp$ and $\tauhm$, \ie the momentum of the system of $\Pgppm$ and $\Pgg$ produced in the decays of the $\Pgtp$ and $\Pgtm$, and $m_{\h^{+}}$ and $m_{\h^{-}}$ denote the masses of these systems.
The symbol $s$ denotes the square of the center-of-mass energy of $10.579$~\GeV and $\epsilon^{\mu\nu\rho\sigma}$ the Levi-Civita tensor.
The symbols $x$, $y$, and $z$ are defined by:
\begin{equation}
x = p_{\Pgt\Pgt} \cdot p_{\h^{+}} \, , 
\quad y = p_{\Pgt\Pgt} \cdot p_{\h^{-}} \, , \quad \textrm{and}
\quad z = p_{\h^{+}} \cdot p_{\h^{-}} \, .
\end{equation}
The fourth coefficient, $d$, is given by:
\begin{equation}
d^{2} = -\frac{1}{4 \, q^{2}} \left[ (1 + a^{2}) \, s + b \, m_{\h^{+}}^{2} + c \, m_{\h^{-}}^{2} - 4 m_{\Pgt}^{2} + 2 \, (a \, c \, y - a \, b \, x - b \, c \, z) \right] \, .
\label{eq:step1_d}
\end{equation}
Eq.~(\ref{eq:step1_d}) yields two solutions of opposite sign, which determine the four-vectors $p_{\Pgtp}$ and $p_{\Pgtm}$ up to a twofold sign ambiguity. We resolve the sign ambiguity by choosing the solution more compatible with tracking information~\cite{Kuhn:1993ra}.
For the decay channels $\Pgppm$ and $\Pgrpm$ we quantify the level of compatibility based on the transverse impact parameter of the charged pion's track, while for the decay channel $\Papm$ we use the position of the $\Pgt$ decay vertex.

\section{Comparison of different methods for measuring spin correlation}
\label{sec:appendixB}

Alternatively to the ML fit given by Eq.~(\ref{eq:lf}), the polarization vectors $\Bvecp$ and $\Bvecm$ for the $\Pgtp$ and $\Pgtm$ and the spin correlation matrix $\CC$ can be measured by:
\begin{itemize}
\item \textbf{Expectation value} \\
Ref.~\cite{Altakach:2022ywa} uses the expectation values of the product of the polarimeter vectors $\hvecp$ and $\hvecm$ to measure the elements of $\Bvecp$, $\Bvecm$, and $\CC$. The relation between the expectation values of $\hvec^{\pm}$ and $\Bvec^{\pm}$ and between $\expval{\hvecp \cdot \hvecm}$ and $\CC$ is given by Eq.~(30) of Ref.~\cite{Altakach:2022ywa}. It reads:
\begin{align}
\BB_i^\pm &= 3 \, \ev{{\bf h}^\pm_{i}} \nonumber \\
\CC_{ij} &= 9 \, \ev{\hvecp_{i} \, \hvecm_{j}} \, ,
\label{eq:exp}
\end{align}
where the indices $i$ and $j$ are either $n$, $r$, or $k$ and the expectation value is computed as average over the events in the $\Pep\Pem \to \Pgtp\Pgtm$ event sample.
\item \textbf{Double-differential cross section} \\
Expressing the Lorentz invariant phase-space measure in Eq.~(\ref{eq:diffXS3}) in polar coordinates and integrating over the azimuthal angles $\phip$ and $\phim$ yields the following expression for the double-differential (2d) cross section as function of the polar angles $\thetap$ and $\thetam$, given by Eq.~(VI.6) in Ref.~\cite{Bernreuther:2004jv}:
\begin{equation}
\frac{1}{\sigma} \, \frac{d\sigma}{d\cos\thetap_{i} \, d\cos\thetam_{j}} = \frac{1}{4} \, \left( 1 + \CC_{ij} \, \cos\thetap_{i} \, \cos\thetam_{j} \right) \, ,
\label{eq:xsec2d}
\end{equation}
where $\cos\thetap_{i} = \hvecp \cdot \hat{e}_{i}$ ($\cos\thetap_{j} = \hvecp \cdot \hat{e}_{j}$) denotes the direction cosine of the polarimetric vector $\hvecp$ ($\hvecm$) with one of the basis vectors $\{\hn, \hr, \hk\}$ in the rest frame of the $\Pgtp$ ($\Pgtm$)
and $i,j\in\{n,r,k\}$.
\item \textbf{Single-differential cross section} \\
The spin correlation matrix $\CC$ may alternatively be extracted from the single-differential (1d) cross section as function of the observable $\xi_{ij} = \cos\thetap_{i} \, \cos\thetam_{j}$, given by Eq.~(4.16) of Ref.~\cite{Bernreuther:2015yna}:
\begin{equation}
\frac{1}{\sigma} \, \frac{d\sigma}{d\xi_{ij}} = \frac{1}{2} \, \left( 1 + \CC_{ij} \, \xi_{ij} \right) \, \ln\left(\frac{1}{\vert\xi\vert}\right) \, .
\label{eq:xsec1d}
\end{equation}
\item \textbf{Forward/backward asymmetry} \\
Alternatively, one may extract the $\Pgt$ spin correlation using the forward--backward (FB) asymmetries given by Eq.~(25) of Ref.~\cite{Aguilar-Saavedra:2022uye}:
\begin{equation}
A_{ij} = \frac{N(\cos\thetap_{i} \, \cos\thetam_{j} > 0) - N(\cos\thetap_{i} \, \cos\thetam_{j} < 0)}{N(\cos\thetap_{i} \, \cos\thetam_{j} > 0) + N(\cos\thetap_{i} \, \cos\thetam_{j} < 0)} = \frac{1}{4} \, \CC_{ij} \, ,
\label{eq:fb}
\end{equation}
where the symbol $N$ represents number of events, the direction cosines $\cos\thetap_{i}$ and $\cos\thetam_{j}$ are defined as before, and $i,j\in\{n,r,k\}$.
\end{itemize}
In case of the double-differential (single-differential) cross section, binned distributions in $\thetap_{i}$ versus $\thetam_{j}$ ($\xi_{ij}$) are fitted to determine the element $\CC_{ij}$ of the spin correlation matrix. The fits are implemented using the software package ROOFIT~\cite{Verkerke:2003ir}.

Different conventions exist in the literature for defining the helicity frame and the sign of the polarimeter vector.
These conventions lead to different signs for the terms proportional to $\BB_i^\pm$ and $\CC_{ij}$ in Eqs.~(\ref{eq:exp}) to~(\ref{eq:fb}).
The signs in the equations above match our definition of the helicity frame and of the polarimeter vector. 

The sensitivity of the different methods is compared in Table~\ref{tab:entanglObs_pipi_mctruth_enhanced}. The ML-fit method provides the lowest uncertainties and thus the highest significance. While the performance of the fits to binned 2d and 1d cross sections comes close to the performance of the ML-fit method, the significances for the expectation value and forward/backward asymmetry methods are about $15\%$ and $30\%$ lower.
The performance advantage of the unbinned ML fit increases if the size of the event sample is reduced.

\begin{table}[ht!]
\centering
\begin{tabular}{l|r@{$ \,\,\pm\,\, $}rr@{$ \,\,\pm\,\, $}rr@{$ \,\,\pm\,\, $}rr@{$ \,\,\pm\,\, $}r}
Method & \multicolumn{2}{c}{$\conc{\rho}$} & \multicolumn{2}{c}{$\Rchsh\qty[\CC]$} \\
\hline
Exp. value & $0.6952$ & $0.0013$ & $1.4270$ & $0.0037$ \\
2d distr.  & $0.6950$ & $0.0013$ & $1.4331$ & $0.0032$ \\
1d distr.  & $0.6949$ & $0.0012$ & $1.4253$ & $0.0033$ \\
FB asymm.  & $0.6932$ & $0.0017$ & $1.4315$ & $0.0045$ \\
ML fit     & $0.6952$ & $0.0011$ & $1.4283$ & $0.0031$ \\
\end{tabular}
\caption{
  Observables $\conc{\rho}$ and $\Rchsh\qty[\CC]$ measured in the combination of decay channels $\Pgpp\Pgpm$, $\Pgppm\Pgrmp$, $\Pgppm\Pamp$, $\Pgrp\Pgrm$, $\Pgrpm\Pamp$, and $\Pap\Pam$,
  for events that pass the selection $\vert\cos(\vartheta)\vert < \cosThetaCut$.
  Events are analyzed on MC-truth level.
  No acceptance cuts are applied on the $\Pgppm$ and $\Pgg$ produced in the $\Pgt$ decays.
}
\label{tab:entanglObs_pipi_mctruth_enhanced}
\end{table}

\twocolumn  

\bibliographystyle{utphys}  
\bibliography{bell_at_belle.bib} 

\providecommand{\href}[2]{#2}\begingroup\raggedright\begin{thebibliography}{10}

\bibitem{Bell:1964kc}
J.~S. Bell, ``{On the Einstein-Podolsky-Rosen paradox},''
  \href{https://dx.doi.org/10.1103/PhysicsPhysiqueFizika.1.195}{{\em Physics
  Physique Fizika} {\bfseries 1} (1964) 195}.

\bibitem{Brunner:RevModPhys.86.419}
N.~Brunner, D.~Cavalcanti, S.~Pironio, V.~Scarani, and S.~Wehner, ``Bell
  nonlocality,'' \href{https://dx.doi.org/10.1103/RevModPhys.86.419}{{\em Rev.
  Mod. Phys.} {\bfseries 86} (Apr, 2014) 419}.
  \url{https://link.aps.org/doi/10.1103/RevModPhys.86.419}.

\bibitem{Aspect:1982fx}
A.~Aspect, J.~Dalibard, and G.~Roger, ``{Experimental test of Bell's
  inequalities using time varying analyzers},''
  \href{https://dx.doi.org/10.1103/PhysRevLett.49.1804}{{\em Phys. Rev. Lett.}
  {\bfseries 49} (1982) 1804}.

\bibitem{Weihs:1998gy}
G.~Weihs, T.~Jennewein, C.~Simon, H.~Weinfurter, and A.~Zeilinger, ``{Violation
  of Bell's inequality under strict Einstein locality conditions},''
  \href{https://dx.doi.org/10.1103/PhysRevLett.81.5039}{{\em Phys. Rev. Lett.}
  {\bfseries 81} (1998) 5039},
  \href{https://arxiv.org/abs/quant-ph/9810080}{{\ttfamily
  arXiv:quant-ph/9810080}}.

\bibitem{Clauser:1969ny}
J.~F. Clauser, M.~A. Horne, A.~Shimony, and R.~A. Holt, ``{Proposed experiment
  to test local hidden variable theories},''
  \href{https://dx.doi.org/10.1103/PhysRevLett.23.880}{{\em Phys. Rev. Lett.}
  {\bfseries 23} (1969) 880}.

\bibitem{Clauser:1974tg}
J.~F. Clauser and M.~A. Horne, ``{Experimental consequences of objective local
  theories},'' \href{https://dx.doi.org/10.1103/PhysRevD.10.526}{{\em Phys.
  Rev. D} {\bfseries 10} (1974) 526}.

\bibitem{Hensen:2015ccp}
B.~Hensen {\em et~al.}, ``{Loophole-free Bell inequality violation using
  electron spins separated by 1.3 kilometres},''
  \href{https://dx.doi.org/10.1038/nature15759}{{\em Nature} {\bfseries 526}
  (2015) 682}, \href{https://arxiv.org/abs/1508.05949}{{\ttfamily
  arXiv:1508.05949 [quant-ph]}}.

\bibitem{Giustina:2015yza}
M.~Giustina {\em et~al.}, ``{Significant-loophole-free test of
  Bell\textquoteright{}s theorem with entangled photons},''
  \href{https://dx.doi.org/10.1103/PhysRevLett.115.250401}{{\em Phys. Rev.
  Lett.} {\bfseries 115} (2015) 250401},
  \href{https://arxiv.org/abs/1511.03190}{{\ttfamily arXiv:1511.03190
  [quant-ph]}}.

\bibitem{Storz:2023jjx}
S.~Storz {\em et~al.}, ``{Loophole-free Bell inequality violation with
  superconducting circuits},''
  \href{https://dx.doi.org/10.1038/s41586-023-05885-0}{{\em Nature} {\bfseries
  617} (2023) 265}.

\bibitem{PhysRevLett.119.010402}
W.~Rosenfeld, D.~Burchardt, R.~Garthoff, K.~Redeker, N.~Ortegel, M.~Rau, and
  H.~Weinfurter, ``{Event-ready Bell test using entangled atoms simultaneously
  closing detection and locality loopholes},''
  \href{https://dx.doi.org/10.1103/PhysRevLett.119.010402}{{\em Phys. Rev.
  Lett.} {\bfseries 119} (Jul, 2017) 010402}.
  \url{https://link.aps.org/doi/10.1103/PhysRevLett.119.010402}.

\bibitem{Clauser:1978ng}
J.~F. Clauser and A.~Shimony, ``{Bell's theorem: experimental tests and
  implications},'' \href{https://dx.doi.org/10.1088/0034-4885/41/12/002}{{\em
  Rept. Prog. Phys.} {\bfseries 41} (1978) 1881}.

\bibitem{Genovese:2005nw}
M.~Genovese, ``{Research on hidden variable theories: a review of recent
  progresses},'' \href{https://dx.doi.org/10.1016/j.physrep.2005.03.003}{{\em
  Phys. Rept.} {\bfseries 413} (2005) 319},
  \href{https://arxiv.org/abs/quant-ph/0701071}{{\ttfamily
  arXiv:quant-ph/0701071}}.

\bibitem{Lamehi-Rachti:1976wey}
M.~Lamehi-Rachti and W.~Mittig, ``{Quantum mechanics and hidden variables: a
  test of Bell's inequality by the measurement of the spin correlation in
  low-energy proton-- proton scattering},''
  \href{https://dx.doi.org/10.1103/PhysRevD.14.2543}{{\em Phys. Rev. D}
  {\bfseries 14} (1976) 2543}.

\bibitem{Tornqvist:1980af}
N.~A. Tornqvist, ``{Suggestion for Einstein-Podolsky-Rosen experiments using
  reactions like $\Pep\Pem \to \Lambda \bar{\Lambda} \to
  \Pgpm\Pp\Pgpp\bar{\Pp}$},'' \href{https://dx.doi.org/10.1007/BF00715204}{{\em
  Found. Phys.} {\bfseries 11} (1981) 171}.

\bibitem{Abel:1992kz}
S.~A. Abel, M.~Dittmar, and H.~K. Dreiner, ``{Testing locality at colliders via
  Bell's inequality?},''
  \href{https://dx.doi.org/10.1016/0370-2693(92)90071-B}{{\em Phys. Lett. B}
  {\bfseries 280} (1992) 304}.

\bibitem{PhysRevD.57.R1332}
F.~Benatti and R.~Floreanini, ``Bell's locality and
  ${\mathrm{\ensuremath{\varepsilon}}}^{\ensuremath{'}}$/\ensuremath{\varepsilon},''
  \href{https://dx.doi.org/10.1103/PhysRevD.57.R1332}{{\em Phys. Rev. D}
  {\bfseries 57} (Feb, 1998) R1332--R1336},
  \href{https://arxiv.org/abs/hep-ph/9712274}{{\ttfamily hep-ph/9712274}}.

\bibitem{Benatti2000}
F.~Benatti and R.~Floreanini, ``{CP-violation as a test of quantum
  mechanics},'' \href{https://dx.doi.org/doi.org/10.1007/s100520000306}{{\em
  Eur. Phys. J. C} {\bfseries 13} (2000) 267},
  \href{https://arxiv.org/abs/hep-ph/9912348}{{\ttfamily
  arXiv:hep-ph/9912348}}.

\bibitem{Bertlmann:2001ea}
R.~A. Bertlmann, W.~Grimus, and B.~C. Hiesmayr, ``{Bell inequality and $CP$
  violation in the neutral kaon system},''
  \href{https://dx.doi.org/10.1016/S0375-9601(01)00577-1}{{\em Phys. Lett. A}
  {\bfseries 289} (2001) 21},
  \href{https://arxiv.org/abs/quant-ph/0107022}{{\ttfamily
  arXiv:quant-ph/0107022}}.

\bibitem{Banerjee:2014vga}
S.~Banerjee, A.~K. Alok, and R.~MacKenzie, ``{Quantum correlations in $\PB$ and
  $\PK$ meson systems},''
  \href{https://dx.doi.org/10.1140/epjp/i2016-16129-0}{{\em Eur. Phys. J. Plus}
  {\bfseries 131} (2016) 129},
  \href{https://arxiv.org/abs/1409.1034}{{\ttfamily arXiv:1409.1034 [hep-ph]}}.

\bibitem{Acin:2000cs}
A.~Acin, J.~I. Latorre, and P.~Pascual, ``{Three party entanglement from
  positronium},'' \href{https://dx.doi.org/10.1103/PhysRevA.63.042107}{{\em
  Phys. Rev. A} {\bfseries 63} (2001) 042107},
  \href{https://arxiv.org/abs/quant-ph/0007080}{{\ttfamily
  arXiv:quant-ph/0007080}}.

\bibitem{Go:2003tx}
{\bfseries Belle} Collaboration, A.~Go, ``{Observation of Bell inequality
  violation in $\PB$ mesons},''
  \href{https://dx.doi.org/10.1080/09500340408233614}{{\em J. Mod. Opt.}
  {\bfseries 51} (2004) 991},
  \href{https://arxiv.org/abs/quant-ph/0310192}{{\ttfamily
  arXiv:quant-ph/0310192}}.

\bibitem{Baranov:2008zzb}
S.~P. Baranov, ``{Bell's inequality in charmonium decays $\eta_{c} \to \Lambda
  \bar{\Lambda}$, $\chi_{c} \to \Lambda\bar{\Lambda}$ and $\textrm{J}/\psi \to
  \Lambda\bar{\Lambda}$},''
  \href{https://dx.doi.org/10.1088/0954-3899/35/7/075002}{{\em J. Phys. G}
  {\bfseries 35} (2008) 075002}.

\bibitem{Banerjee:2015mha}
S.~Banerjee, A.~K. Alok, R.~Srikanth, and B.~C. Hiesmayr, ``{A quantum
  information theoretic analysis of three flavor neutrino oscillations},''
  \href{https://dx.doi.org/10.1140/epjc/s10052-015-3717-x}{{\em Eur. Phys. J.
  C} {\bfseries 75} no.~10, (2015) 487},
  \href{https://arxiv.org/abs/1508.03480}{{\ttfamily arXiv:1508.03480
  [hep-ph]}}.

\bibitem{Yongram:2013soa}
N.~Yongram and E.~B. Manoukian, ``{Quantum field theory analysis of
  polarizations correlations, entanglement and Bell's inequality: explicit
  processes},'' \href{https://dx.doi.org/10.1002/prop.201200137}{{\em Fortsch.
  Phys.} {\bfseries 61} (2013) 668},
  \href{https://arxiv.org/abs/1309.2059}{{\ttfamily arXiv:1309.2059 [hep-th]}}.

\bibitem{Cervera-Lierta:2017tdt}
A.~Cervera-Lierta, J.~I. Latorre, J.~Rojo, and L.~Rottoli, ``{Maximal
  entanglement in high energy physics},''
  \href{https://dx.doi.org/10.21468/SciPostPhys.3.5.036}{{\em SciPost Phys.}
  {\bfseries 3} (2017) 036}, \href{https://arxiv.org/abs/1703.02989}{{\ttfamily
  arXiv:1703.02989 [hep-th]}}.

\bibitem{Afik:2020onf}
Y.~Afik and J.~R.~M. de~Nova, ``{Entanglement and quantum tomography with top
  quarks at the LHC},''
  \href{https://dx.doi.org/10.1140/epjp/s13360-021-01902-1}{{\em Eur. Phys. J.
  Plus} {\bfseries 136} (2021) 907},
  \href{https://arxiv.org/abs/2003.02280}{{\ttfamily arXiv:2003.02280
  [quant-ph]}}.

\bibitem{Fabbrichesi:2021npl}
M.~Fabbrichesi, R.~Floreanini, and G.~Panizzo, ``{Testing Bell inequalities at
  the LHC with top-quark pairs},''
  \href{https://dx.doi.org/10.1103/PhysRevLett.127.161801}{{\em Phys. Rev.
  Lett.} {\bfseries 127} (2021) 161801},
  \href{https://arxiv.org/abs/2102.11883}{{\ttfamily arXiv:2102.11883
  [hep-ph]}}.

\bibitem{Severi:2021cnj}
C.~Severi, C.~D.~E. Boschi, F.~Maltoni, and M.~Sioli, ``{Quantum tops at the
  LHC: from entanglement to Bell inequalities},''
  \href{https://dx.doi.org/10.1140/epjc/s10052-022-10245-9}{{\em Eur. Phys. J.
  C} {\bfseries 82} (2022) 285},
  \href{https://arxiv.org/abs/2110.10112}{{\ttfamily arXiv:2110.10112
  [hep-ph]}}.

\bibitem{Larkoski:2022lmv}
A.~J. Larkoski, ``{General analysis for observing quantum interference at
  colliders},'' \href{https://dx.doi.org/10.1103/PhysRevD.105.096012}{{\em
  Phys. Rev. D} {\bfseries 105} (2022) 096012},
  \href{https://arxiv.org/abs/2201.03159}{{\ttfamily arXiv:2201.03159
  [hep-ph]}}.

\bibitem{Aguilar-Saavedra:2022uye}
J.~A. Aguilar-Saavedra and J.~A. Casas, ``{Improved tests of entanglement and
  Bell inequalities with LHC tops},''
  \href{https://dx.doi.org/10.1140/epjc/s10052-022-10630-4}{{\em Eur. Phys. J.
  C} {\bfseries 82} (2022) 666},
  \href{https://arxiv.org/abs/2205.00542}{{\ttfamily arXiv:2205.00542
  [hep-ph]}}.

\bibitem{Afik:2022dgh}
Y.~Afik and J.~R.~M. de~Nova, ``{Quantum discord and steering in top quarks at
  the LHC},'' \href{https://dx.doi.org/10.1103/PhysRevLett.130.221801}{{\em
  Phys. Rev. Lett.} {\bfseries 130} (2023) 221801},
  \href{https://arxiv.org/abs/2209.03969}{{\ttfamily arXiv:2209.03969
  [quant-ph]}}.

\bibitem{Afik:2022kwm}
Y.~Afik and J.~R.~M. de~Nova, ``{Quantum information with top quarks in QCD},''
  \href{https://dx.doi.org/10.22331/q-2022-09-29-820}{{\em Quantum} {\bfseries
  6} (2022) 820}, \href{https://arxiv.org/abs/2203.05582}{{\ttfamily
  arXiv:2203.05582 [quant-ph]}}.

\bibitem{Gong:2021bcp}
W.~Gong, G.~Parida, Z.~Tu, and R.~Venugopalan, ``{Measurement of Bell-type
  inequalities and quantum entanglement from $\Lambda$-hyperon spin
  correlations at high energy colliders},''
  \href{https://dx.doi.org/10.1103/PhysRevD.106.L031501}{{\em Phys. Rev. D}
  {\bfseries 106} (2022) L031501},
  \href{https://arxiv.org/abs/2107.13007}{{\ttfamily arXiv:2107.13007
  [hep-ph]}}.

\bibitem{Barr:2021zcp}
A.~J. Barr, ``{Testing Bell inequalities in Higgs boson decays},''
  \href{https://dx.doi.org/10.1016/j.physletb.2021.136866}{{\em Phys. Lett. B}
  {\bfseries 825} (2022) 136866},
  \href{https://arxiv.org/abs/2106.01377}{{\ttfamily arXiv:2106.01377
  [hep-ph]}}.

\bibitem{Aguilar-Saavedra:2022wam}
J.~A. Aguilar-Saavedra, A.~Bernal, J.~A. Casas, and J.~M. Moreno, ``{Testing
  entanglement and Bell inequalities in $\PH \to \PZ\PZ$},''
  \href{https://dx.doi.org/10.1103/PhysRevD.107.016012}{{\em Phys. Rev. D}
  {\bfseries 107} (2023) 016012},
  \href{https://arxiv.org/abs/2209.13441}{{\ttfamily arXiv:2209.13441
  [hep-ph]}}.

\bibitem{Ashby-Pickering:2022umy}
R.~Ashby-Pickering, A.~J. Barr, and A.~Wierzchucka, ``{Quantum state
  tomography, entanglement detection and Bell violation prospects in weak
  decays of massive particles},''
  \href{https://dx.doi.org/10.1007/JHEP05(2023)020}{{\em JHEP} {\bfseries 05}
  (2023) 020}, \href{https://arxiv.org/abs/2209.13990}{{\ttfamily
  arXiv:2209.13990 [quant-ph]}}.

\bibitem{Fabbrichesi:2023cev}
M.~Fabbrichesi, R.~Floreanini, E.~Gabrielli, and L.~Marzola, ``{Bell
  inequalities and quantum entanglement in weak gauge bosons production at the
  LHC and future colliders},''
  \href{https://arxiv.org/abs/2302.00683}{{\ttfamily arXiv:2302.00683
  [hep-ph]}}.

\bibitem{ATLAS:2023fsd}
{\bfseries ATLAS} Collaboration, G.~Aad {\em et~al.}, ``{Observation of quantum
  entanglement in top-quark pairs using the ATLAS detector},''
  \href{https://arxiv.org/abs/2311.07288}{{\ttfamily arXiv:2311.07288
  [hep-ex]}}.

\bibitem{Fabbrichesi:2023idl}
M.~Fabbrichesi, R.~Floreanini, E.~Gabrielli, and L.~Marzola, ``{Bell inequality
  is violated in $\PB^0 \to \textrm{J}/\psi \, \PK^{\ast}(892)^0$ decays},''
  \href{https://arxiv.org/abs/2305.04982}{{\ttfamily arXiv:2305.04982
  [hep-ph]}}.

\bibitem{Belle:2020lfn}
{\bfseries Belle} Collaboration, D.~Sahoo {\em et~al.}, ``{Search for
  lepton-number- and baryon-number-violating tau decays at Belle},''
  \href{https://dx.doi.org/10.1103/PhysRevD.102.111101}{{\em Phys. Rev. D}
  {\bfseries 102} (2020) 111101},
  \href{https://arxiv.org/abs/2010.15361}{{\ttfamily arXiv:2010.15361
  [hep-ex]}}.

\bibitem{Belle-II:2023izd}
{Belle-II Collaboration}, ``{Measurement of the $\Pgt$-lepton mass with the
  Belle II experiment},''
  \href{https://dx.doi.org/10.1103/PhysRevD.108.032006}{{\em Phys. Rev. D}
  {\bfseries 108} (2023) 032006},
  \href{https://arxiv.org/abs/2305.19116}{{\ttfamily arXiv:2305.19116
  [hep-ex]}}.

\bibitem{Akai:2018mbz}
{SuperKEKB Collaboration}, ``{SuperKEKB Collider},''
  \href{https://dx.doi.org/10.1016/j.nima.2018.08.017}{{\em Nucl. Instrum.
  Meth. A} {\bfseries 907} (2018) 188},
  \href{https://arxiv.org/abs/1809.01958}{{\ttfamily arXiv:1809.01958
  [physics.acc-ph]}}.

\bibitem{Belle-II:2018jsg}
{\bfseries Belle-II} Collaboration, W.~Altmannshofer {\em et~al.}, ``{The Belle
  II physics book},'' \href{https://dx.doi.org/10.1093/ptep/ptz106}{{\em PTEP}
  {\bfseries 2019} (2019) 123C01},
  \href{https://arxiv.org/abs/1808.10567}{{\ttfamily arXiv:1808.10567
  [hep-ex]}}. [Erratum: PTEP 2020, 029201 (2020)].

\bibitem{Privitera:1991nz}
P.~Privitera, ``{Decay correlations in $\Pep\Pem \to \Pgtp\Pgtm$ as a test of
  quantum mechanics},''
  \href{https://dx.doi.org/10.1016/0370-2693(92)90872-2}{{\em Phys. Lett. B}
  {\bfseries 275} (1992) 172}.

\bibitem{Fabbrichesi:2022ovb}
M.~Fabbrichesi, R.~Floreanini, and E.~Gabrielli, ``{Constraining new physics in
  entangled two-qubit systems: top-quark, tau-lepton and photon pairs},''
  \href{https://dx.doi.org/10.1140/epjc/s10052-023-11307-2}{{\em Eur. Phys. J.
  C} {\bfseries 83} (2023) 162},
  \href{https://arxiv.org/abs/2208.11723}{{\ttfamily arXiv:2208.11723
  [hep-ph]}}.

\bibitem{Altakach:2022ywa}
M.~M. Altakach, P.~Lamba, F.~Maltoni, K.~Mawatari, and K.~Sakurai, ``{Quantum
  information and CP measurement in $\PH \to \Pgtp\Pgtm$ at future lepton
  colliders},'' \href{https://dx.doi.org/10.1103/PhysRevD.107.093002}{{\em
  Phys. Rev. D} {\bfseries 107} (2023) 093002},
  \href{https://arxiv.org/abs/2211.10513}{{\ttfamily arXiv:2211.10513
  [hep-ph]}}.

\bibitem{Ma:2023yvd}
K.~Ma and T.~Li, ``{Testing Bell inequality through $\PH \to \Pgt\Pgt$ at
  CEPC},'' \href{https://arxiv.org/abs/2309.08103}{{\ttfamily arXiv:2309.08103
  [hep-ph]}}.

\bibitem{Bernreuther:2001rq}
W.~Bernreuther, A.~Brandenburg, Z.~G. Si, and P.~Uwer, ``{Top quark spin
  correlations at hadron colliders: predictions at next-to-leading order
  QCD},'' \href{https://dx.doi.org/10.1103/PhysRevLett.87.242002}{{\em Phys.
  Rev. Lett.} {\bfseries 87} (2001) 242002},
  \href{https://arxiv.org/abs/hep-ph/0107086}{{\ttfamily
  arXiv:hep-ph/0107086}}.

\bibitem{Bouchiat:1958yui}
C.~Bouchiat and L.~Michel, ``{Mesure de la polarisation des electrons
  relativistes},'' \href{https://dx.doi.org/10.1016/0029-5582(58)90046-4}{{\em
  Nucl. Phys.} {\bfseries 5} (1958) 416}.

\bibitem{Bennett:1996gf}
C.~H. Bennett, D.~P. DiVincenzo, J.~A. Smolin, and W.~K. Wootters, ``{Mixed
  state entanglement and quantum error correction},''
  \href{https://dx.doi.org/10.1103/PhysRevA.54.3824}{{\em Phys. Rev. A}
  {\bfseries 54} (1996) 3824},
  \href{https://arxiv.org/abs/quant-ph/9604024}{{\ttfamily
  arXiv:quant-ph/9604024}}.

\bibitem{Horodecki:2009zz}
R.~Horodecki, P.~Horodecki, M.~Horodecki, and K.~Horodecki, ``{Quantum
  entanglement},'' \href{https://dx.doi.org/10.1103/RevModPhys.81.865}{{\em
  Rev. Mod. Phys.} {\bfseries 81} (2009) 865},
  \href{https://arxiv.org/abs/quant-ph/0702225}{{\ttfamily
  arXiv:quant-ph/0702225}}.

\bibitem{Wootters:1997id}
W.~K. Wootters, ``{Entanglement of formation of an arbitrary state of two
  qubits},'' \href{https://dx.doi.org/10.1103/PhysRevLett.80.2245}{{\em Phys.
  Rev. Lett.} {\bfseries 80} (1998) 2245},
  \href{https://arxiv.org/abs/quant-ph/9709029}{{\ttfamily
  arXiv:quant-ph/9709029}}.

\bibitem{Clauser_1978}
J.~F. Clauser and A.~Shimony, ``Bell's theorem. experimental tests and
  implications,'' \href{https://dx.doi.org/10.1088/0034-4885/41/12/002}{{\em
  Reports on Progress in Physics} {\bfseries 41} (Dec, 1978) 1881}.
  \url{https://dx.doi.org/10.1088/0034-4885/41/12/002}.

\bibitem{HORODECKI1995340}
R.~Horodecki, P.~Horodecki, and M.~Horodecki, ``Violating bell inequality by
  mixed spin-12 states: necessary and sufficient condition,''
  \href{https://dx.doi.org/https://doi.org/10.1016/0375-9601(95)00214-N}{{\em
  Phys. Lett. A} {\bfseries 200} (1995) 340}.
  \url{https://www.sciencedirect.com/science/article/pii/037596019500214N}.

\bibitem{Bell:1987hh}
J.~S. Bell, {\em {Spekable and unspeakable in quantum mechanics. Collected
  papers on quantum philosophy}}.
\newblock Cambridge University Press, 1987.

\bibitem{Jadach:1990mz}
S.~Jadach, J.~H. K{\"u}hn, and Z.~Was, ``{TAUOLA: a library of Monte Carlo
  programs to simulate decays of polarized tau leptons},''
  \href{https://dx.doi.org/10.1016/0010-4655(91)90038-M}{{\em Comput. Phys.
  Commun.} {\bfseries 64} (1990) 275}.

\bibitem{Cherepanov:2023wfp}
V.~Cherepanov and C.~Veelken, ``{The polarimeter vector for $\Pgt \to 3
  \Pgp\Pgngt$ decays},'' \href{https://arxiv.org/abs/2311.10490}{{\ttfamily
  arXiv:2311.10490 [hep-ex]}}.

\bibitem{Kuhn:1995nn}
J.~H. K{\"u}hn, ``{Tau polarimetry with multi-meson states},''
  \href{https://dx.doi.org/10.1103/PhysRevD.52.3128}{{\em Phys. Rev. D}
  {\bfseries 52} (1995) 3128},
  \href{https://arxiv.org/abs/hep-ph/9505303}{{\ttfamily
  arXiv:hep-ph/9505303}}.

\bibitem{Davier:1992nw}
M.~Davier, L.~Duflot, F.~Le~Diberder, and A.~Rouge, ``{The optimal method for
  the measurement of tau polarization},''
  \href{https://dx.doi.org/10.1016/0370-2693(93)90101-M}{{\em Phys. Lett. B}
  {\bfseries 306} (1993) 411}.

\bibitem{Rossi:2018}
R.~J. Rossi, {\em {Mathematical statistics: an introduction to likelihood based
  inference}}.
\newblock {John Wiley \& Sons}, 2018.

\bibitem{James:1975dr}
F.~James and M.~Roos, ``{Minuit: a system for function minimization and
  analysis of the parameter errors and correlations},''
  \href{https://dx.doi.org/10.1016/0010-4655(75)90039-9}{{\em Comput. Phys.
  Commun.} {\bfseries 10} (1975) 343}.

\bibitem{Alwall:2014hca}
J.~Alwall, R.~Frederix, S.~Frixione, V.~Hirschi, F.~Maltoni, O.~Mattelaer,
  H.~S. Shao, T.~Stelzer, P.~Torrielli, and M.~Zaro, ``{The automated
  computation of tree-level and next-to-leading order differential cross
  sections, and their matching to parton shower simulations},''
  \href{https://dx.doi.org/10.1007/JHEP07(2014)079}{{\em JHEP} {\bfseries 07}
  (2014) 079}, \href{https://arxiv.org/abs/1405.0301}{{\ttfamily
  arXiv:1405.0301 [hep-ph]}}.

\bibitem{Bierlich:2022pfr}
C.~Bierlich {\em et~al.}, ``{A comprehensive guide to the physics and usage of
  PYTHIA 8.3}'' \href{https://arxiv.org/abs/2203.11601}{{\ttfamily
  arXiv:2203.11601 [hep-ph]}}.

\bibitem{Belle-II:2010dht}
{Belle-II Collaboration}, ``{Belle II technical design report},''
  \href{https://arxiv.org/abs/1011.0352}{{\ttfamily arXiv:1011.0352
  [physics.ins-det]}}.

\bibitem{GEANT4:2002zbu}
{GEANT4 Collaboration}, ``{GEANT4--a simulation toolkit},''
  \href{https://dx.doi.org/10.1016/S0168-9002(03)01368-8}{{\em Nucl. Instrum.
  Meth. A} {\bfseries 506} (2003) 250}.

\bibitem{Avery:kinfit}
P.~Avery, ``{Fitting theory writeups and references},''
  {https://www.phys.ufl.edu/~avery/fitting.html}.

\bibitem{Sauerland:1358627}
P.~Sauerland, {\em {Kinematic reconstruction of tau leptons and test for lepton
  universality in charged weak interactions with the CMS experiment}}.
\newblock PhD thesis, {RWTH Aachen}, 2011.
\newblock \url{{https://cds.cern.ch/record/1358627}}.

\bibitem{Cherepanov:2018npf}
V.~Cherepanov and A.~Zotz, ``{Kinematic reconstruction of $\PZ/\PH \to
  \Pgt\Pgt$ decay in proton-proton collisions},''
  \href{https://arxiv.org/abs/1805.06988}{{\ttfamily arXiv:1805.06988
  [hep-ph]}}.

\bibitem{Avery:hugeerror}
P.~Avery, ``{Applied fitting theory I -- general least squares theory},''
  {https://www.phys.ufl.edu/~avery/fitting/fitting1.pdf}.

\bibitem{bootstrapping}
B.~Efron, ``{Bootstrap methods: another look at the Jackknife},''
  \href{https://dx.doi.org/10.1214/aos/1176344552}{{\em Ann. Statist.}
  {\bfseries 7} (1979) 1}.

\bibitem{Kuhn:1993ra}
J.~H. K{\"u}hn, ``{Tau kinematics from impact parameters},''
  \href{https://dx.doi.org/10.1016/0370-2693(93)90019-E}{{\em Phys. Lett. B}
  {\bfseries 313} (1993) 458},
  \href{https://arxiv.org/abs/hep-ph/9307269}{{\ttfamily
  arXiv:hep-ph/9307269}}.

\bibitem{Bernreuther:2004jv}
W.~Bernreuther, A.~Brandenburg, Z.~G. Si, and P.~Uwer, ``{Top quark pair
  production and decay at hadron colliders},''
  \href{https://dx.doi.org/10.1016/j.nuclphysb.2004.04.019}{{\em Nucl. Phys. B}
  {\bfseries 690} (2004) 81},
  \href{https://arxiv.org/abs/hep-ph/0403035}{{\ttfamily
  arXiv:hep-ph/0403035}}.

\bibitem{Bernreuther:2015yna}
W.~Bernreuther, D.~Heisler, and Z.-G. Si, ``{A set of top quark spin
  correlation and polarization observables for the LHC: Standard Model
  predictions and new physics contributions},''
  \href{https://dx.doi.org/10.1007/JHEP12(2015)026}{{\em JHEP} {\bfseries 12}
  (2015) 026}, \href{https://arxiv.org/abs/1508.05271}{{\ttfamily
  arXiv:1508.05271 [hep-ph]}}.

\bibitem{Verkerke:2003ir}
W.~Verkerke and D.~P. Kirkby, ``{The RooFit toolkit for data modeling},'' {\em
  eConf} {\bfseries C0303241} (2003) MOLT007,
  \href{https://arxiv.org/abs/physics/0306116}{{\ttfamily
  arXiv:physics/0306116}}.

\end{thebibliography}\endgroup
\end{document}